\documentclass[Afour,sageh,times]{sagej}

\usepackage{moreverb,url}

\usepackage[colorlinks,bookmarksopen,bookmarksnumbered,citecolor=red,urlcolor=red]{hyperref}
\usepackage{listings}
\usepackage{xcolor}

\definecolor{codegreen}{rgb}{0,0.6,0}
\definecolor{codegray}{rgb}{0.5,0.5,0.5}
\definecolor{codepurple}{rgb}{0.58,0,0.82}
\definecolor{backcolour}{rgb}{0.95,0.95,0.92}

\lstdefinestyle{mystyle}{
    backgroundcolor=\color{backcolour},   
    commentstyle=\color{codegreen},
    keywordstyle=\color{magenta},
    numberstyle=\tiny\color{codegray},
    stringstyle=\color{codepurple},
    basicstyle=\ttfamily\footnotesize,
    breakatwhitespace=false,         
    breaklines=true,                 
    captionpos=b,                    
    keepspaces=true,                 
    numbers=left,                    
    numbersep=5pt,                  
    showspaces=false,                
    showstringspaces=false,
    showtabs=false,                  
    tabsize=2
}

\newcommand{\commentout}[1]{}

\newcommand{\OLD}[1]{{{\color{blue}#1}}}

\newcommand{\REPORT}[1]{{{\color{cyan}#1}}}

\newcommand\BibTeX{{\rmfamily B\kern-.05em \textsc{i\kern-.025em b}\kern-.08em
T\kern-.1667em\lower.7ex\hbox{E}\kern-.125emX}}

\def\volumeyear{2020}

\begin{document}

\runninghead{Zhang \textit{et~al.}}

\title{AMReX: Block-Structured Adaptive Mesh Refinement for Multiphysics Applications}

\author{Weiqun Zhang\affilnum{1},  Andrew Myers\affilnum{1}, Kevin Gott \affilnum{2}, Ann Almgren\affilnum{1}, John Bell\affilnum{1}}

\affiliation{\affilnum{1}CCSE, Lawrence Berkeley National Laboratory, Berkeley, CA, USA\\
\affilnum{2}NERSC, Lawrence Berkeley National Laboratory, Berkeley, CA, USA}
%\affilnum{2}SAGE Publications Ltd, UK}

\corrauth{Weiqun Zhang, MS 50A-3111, Lawrence Berkeley, National Laboratory, Berkeley, CA   94720.
}

\email{weiqunzhang@lbl.gov}

\begin{abstract}
Block-structured adaptive mesh refinement (AMR) provides the basis for the temporal and spatial discretization strategy for a number of ECP applications in the areas of accelerator design, additive manufacturing, astrophysics, combustion, cosmology, multiphase flow,  and wind plant modelling.  AMReX is a software framework that provides a unified infrastructure with the functionality needed for these and other AMR applications to be able to effectively and efficiently utilize machines from laptops to exascale architectures. AMR reduces the computational cost and memory footprint compared to a uniform mesh while preserving accurate descriptions of different physical processes in complex multi-physics algorithms. AMReX supports algorithms that solve systems of partial differential equations (PDEs) in simple or complex geometries, and those that use particles and/or particle-mesh operations to represent component physical processes.  In this paper, we will discuss the core elements of the AMReX framework such as data containers and iterators as well as several specialized operations to meet the needs of the application projects.  In addition we will highlight the strategy that the AMReX team is pursuing to achieve highly performant code across a range of accelerator-based architectures for a variety of different applications.
\end{abstract}


\maketitle

%\INFO{
%\section{INSTRUCTIONS}
%This special journal issue will be comprised of contributions from the ECP Co-design center principal investigators and their teams, in which they describe the motivations for creation of each center, their implementation of a co-design paradigm in the preparation of exascale-capable applications and hardware, and achievements to date. Things for inclusion:

%    Case studies, where relevant
%    Engagement with vendors and software projects
%    Performance/portability strategies and results %}

\section{Introduction}
AMReX, a software framework developed and supported by the U.S. DOE Exascale Computing Project's %h\MarginPar{Project or Program?} 
AMReX Co-Design Center, supports the development of block-structured adaptive mesh refinement (AMR) algorithms for solving systems of partial differential equations, in simple or complex geometries, on machines from laptops to exascale architectures. Block-structured AMR provides the basis for the temporal and spatial discretization strategy for ECP applications in
accelerator design, additive manufacturing, astrophysics, combustion, cosmology, multiphase flow and wind energy. 
This suite of applications represents a broad range of different physical processes with a wide range of algorithmic requirements.

%Many applications use particles to represent some of the physical processes.  Some applications need to represent a complex flow domain.
%Some applications use purely explicit discretization while others include implicit components.
%AMReX provides a unified infrastructure with the functionality needed for these and other AMR applications to be able to effectively utilize exascale architectures. AMR reduces the computational cost and memory footprint compared to a uniform mesh while preserving the local descriptions of different physical processes in complex multi-physics algorithms. 

Fundamental to block-structured AMR algorithms is a hierarchical representation of the solution at multiple levels of resolution. At each level the solution is defined on the union of data containers at that resolution, each of which represents the solution over a logically rectangular subregion of the domain. For mesh-based algorithms, AMReX provides support for both explicit and implicit discretizations.  Support for solving linear systems includes native geometric multigrid solvers as well the ability to link to external linear algebra software.  For problems that include stiff systems of ordinary differential equations that represent single-point processes such as chemical kinetics or nucleosynthesis, AMReX provides an interface to ODE solvers provided by SUNDIALS.  AMReX supports algorithms that utilize particles and/or particle-mesh operations to represent component physical processes.
AMReX also provides support for embedded boundary (cut cell) representations of complex geometries.  Both particles and embedded boundary representations introduce additional irregularity and complexity in the way data is stored and operated on, requiring special attention in the presence of dynamically changing hierarchical mesh structures and AMR time stepping approaches.

There are a number of different open-source AMR frameworks; see the 2014 survey paper by \cite{DubeyETAL:2014} for those available at the time.  Most of these frameworks originally used MPI for basic parallelization where data is distributed to MPI ranks with each rank performing operations on its own data.  More recently, most of these frameworks, including BoxLib, the precursor to AMReX, introduced some level of hierarchical parallelism through the use of OpenMP directives.  However, as we have been moving towards the exascale era, node architectures have been changing dramatically, with the advent and increasingly widespread use of many-core CPU-only nodes and hybrid nodes with GPU accelerators.  Further complicating the landscape is the fact that different types of GPUs have different programming models.  A key driver behind the development of AMReX is to provide a high performance framework for applications to run on a variety of current- and next-generation systems without substantial architecture-specific code development / modification costs for each new platform.

In the remainder of the paper, we first define what we mean by block-structured AMR followed by a discussion of the basic design issues considered in the development of AMReX.  We then discuss the ECP applications that are using AMReX and briefly describe their computational requirements.   Next we discuss the data structures used to describe the grid hierarchy and the data containers used to hold mesh data.  We then discuss additional features of AMReX needed to support particle-based algorithms and embedded boundary discretizations.  We also discuss linear solvers needed for implicit discretizations.  We close with a brief discussion of software engineering practice and some concluding remarks.

%Solution strategies vary from level-by-level approaches (with or without subcycling in time) with multilevel synchronization to full-hierarchy approaches, and any combination thereof. AMReX provides data containers and iterators that understand the underlying hierarchical parallelism for field variables on a mesh, particle data and embedded boundary (cut cell) representations of complex geometries.   AMReX also provides performance portability, enabling AMReX-based applications to move between CPU-based architectures and different hybrid CPU-accelerator systems with minimal changes to the application code itself.

%Additional text at
%https://www.exascaleproject.org/amrex-co-design-center-helps-five-application-projects-reach-performance-goals

\section{Block-Structured AMR}

In the context of this paper, block-structured AMR is considered to have the following defining features:
\begin{itemize}
    \item The mesh covering the computational domain is decomposed spatially into structured patches (grids) that each cover a logically rectangular region of the domain.
    \item Patches with the same mesh spacing are disjoint; the union of such patches is referred to as a level.  Only the coarsest level ($\ell = 0$) is required to cover the domain, though finer levels can cover it as well.
    \item The complete mesh hierarchy on which field variables are defined is the union of all the levels.  Proper nesting is enforced, i.e. the union of grids at level $\ell > 0$ is strictly contained within the union of grids at level $\ell-1.$  We note, however, that AMReX does not impose proper nesting at the individual grid level; i.e., a level $\ell$ grid can span multiple grids at level $\ell-1$.
    \item The decomposition of the physical region covered by each level can be different for mesh data vs. particle data; we refer to this as a "dual grid" approach
    \commentout{be decomposed into different patches to support particle vs mesh data.}
    \item The mesh hierarchy can change dynamically throughout a simulation.
\end{itemize}

We note the contrast to block-structured AMR frameworks that use a quadtree/octree structure, such as FLASH \citep{Fryxell2000,Dubey2013}, where the patches have a uniform size, each grid at level $\ell > 0$ has a unique parent grid at level $\ell-1,$ and algorithms are typically constrained to use the same time step at all levels.  We note that these grid patterns and algorithms are a subset of the AMReX capabilities.

%\section{Philosophy of Performance Portability (1 page)}
\section{Design Philosophy}
There are two major factors that inform the design of the AMReX framework.
First, AMReX should be able to support a  wide  range  of  multiphysics  applications  with  different performance characteristics. 
Second, AMReX should not impose restrictions on how application developers construct their algorithms.
\commentout{A guiding principle for the AMReX design is to maintain flexibility in spatial discretizations and time-stepping strategies.}
Consequently, AMReX must provide a rich set of tools with sufficient flexibility that the software can meet the algorithmic requirements of many different applications without sacrificing the performance of any.
To achieve these goals, careful attention has been paid to separating the design of the data structures and basic operations from the algorithms that use those data structures.
The core software components provide the flexibility to support the exploration, development and implementation of new algorithms that might generate additional performance gains.
 
The AMReX design allows application developers to interact with the software at several different levels of abstraction. 
It is possible to simply use the AMReX data containers and iterators and none of the higher-level functionality.  \commentout{In one of the most popular approaches, the developer} 
A more popular approach is to use the data structures and iterators for single- and multi-level operations but retain complete control over the time evolution algorithm, i.e., the ordering of algorithmic components at each level and across levels.  In an alternative approach, the developer exploits additional functionality in AMReX that is designed \commentout{in particular} specifically to support traditional subcycling-in-time algorithms. In this approach, stubs are provided for the necessary operations such as advancing the solution on a level, correcting coarse grid fluxes with time- and space-averaged fine grid fluxes, averaging data from fine to coarse and interpolating in both space and time from coarse to fine.
This layered design provides users with the ability to have complete control over their algorithm or to utilize an application template that can provide higher-level functionality.

The other major factor influencing the AMReX design is performance portability.
As we enter the exascale era, a variety of new architectures are being developed with different
capabilities and programming models. Our goal is to isolate applications from any particular architecture and programming model without sacrificing performance.  To achieve this goal, we introduced a light weight abstraction layer that effectively hides the details of the architecture from the application.  
\commentout{The basic idea here is to introduce}
This layer provides constructs that allow the user to specify what operations they want to perform on a block of data without specifying how those operations are carried out. AMReX then maps those operations onto the hardware at compile time so that the hardware is utilized effectively.
For example, on a many-core node, an operation would be mapped onto a tiled execution model using OpenMP to guarantee good cache performance while 
on a different architecture the same operation \commentout{would} might be mapped to a kernel launch appropriate to a particular GPU.

Similar in spirit to the other aspects of the design, applications are not required to use the 
\commentout{AMReX abstraction layer}
AMReX-provided kernel launch functionality.  Applications can use OpenMP, OpenACC or native programming models to implement their own kernel launches if desired. Furthermore, as part of the AMReX design, specific language requirements are not imposed on users (although \commentout{some} much of the AMReX internal performance portability capability requires C++).
Specifically, the project supports application modules written in Fortran, C, C++ or other languages that can be linked to C++.

\section{AMReX-based ECP Applications}
Seven ECP application projects, 
%—in the areas of accelerator design (WarpX), astrophysics (ExaStar), combustion (Pele), cosmology (ExaSky), multiphaseflow (MFIX-Exa), wind plant modeling (ExaWind) and additive manufacturing (ExaAM) -- 
as well as numerous non-ECP projects, include codes based on AMReX. Here we briefly review the ECP applications and how they use AMReX.
%All codes make use of the basic mesh data structures and iterators along with additional capabilities as discussed below.

%Many applications use particles to represent some of the physical processes.  Some applications need to represent a complex flow domain.
%Some applications use purely explicit discretization while others include implicit components.
%AMReX provides a unified infrastructure with the functionality needed for these and other AMR applications to be able to effectively utilize exascale architectures. 

%\begin{itemize}
WarpX \citep{WarpX} is a multilevel electromagnetic PIC code for simulation of plasma accelerators; electrons are modeled as AMReX particles while the electric and magnetic fields are defined (at edges and faces, respectively) on the hierarchical mesh.  The PIC algorithm heavily leverages the particle-mesh functionality with particle data deposited onto the mesh and mesh data interpolated to particles in every time step.

The Castro \citep{Castro} code for compressible astrophysics is part of the ExaStar project and uses the built-in support for subcycling in time.  The basic time advance includes explicit hydrodynamics,  solution of the Poisson equation for self-gravity, and time integration of stiff nuclear reaction networks. 

Nyx \citep{Nyx} also solves the equations of compressible hydrodynamics with self-gravity, but here coupled with an N-body representation of dark matter in the context of an evolving universe.   Dark matter particles interact with each other and with the mesh-based fluid only through gravitational forcing.   Time integration of the stiff source terms is typically done with a call to CVODE, a solver provided by the SUNDIALS project \citep{hindmarsh2005sundials}, one of the ECP Software Technology projects .

The \cite{MFIX-Exa} multiphase modeling code also heavily leverages both the mesh and particle functionality; the particles represent solid particles within a gas.  Unlike the applications mentioned above, in MFiX-Exa particle / particle interactions play a major role in the dynamics.  Typical MFiX-Exa applications take place inside non-rectangular geometries represented using the EB data structures. Multigrid solvers are used to solve for the dynamic pressure field in a projection formulation and for the implicit treatment of viscous terms.

 The  compressible  combustion  code, \cite{PeleC},  and  the  low  Mach  number  combustion  modeling code, \cite{PeleLM}, are both based on AMReX. Both use the EB methodology to represent the problem geometry, and possibly CVODE to evolve the chemical kinetics.  Like MFiX-Exa, PeleLM uses the multigrid solvers to solve for the dynamic pressure field in a projection formulation and for the semi-implicit treatment of viscous terms.  Particles can be used both as tracer particles and to represent sprays.
 
The ExaWind project combines AMR-Wind (\citeyear{AMR-Wind}), an AMReX-based multilevel structured flow solver with Nalu-Wind, an unstructured flow solver. The flow solvers are coupled using an overset mesh approach handled by the TIOGA library. Both Nalu-Wind and AMR-Wind solve the incompressible Navier-Stokes equations with additional physics to model atmospheric boundary layers. In a wind-farm simulation Nalu-Wind is designed to resolve the complicated geometry and flow near the wind turbine blades while AMR-Wind solves for the flow in the full domain away from the turbines.
 
One  of  the  codes  in  the  ExaAM  project,  TruchasPBF,  is  based  on  AMReX.  TruchasPBF is a finite volume code that models free surface flows with heat transfer, phase change and species diffusion.  TruchasPBF is used for continuum modeling of melt pool physics.
%\end{itemize}

\section{Creating and managing the grid hierarchy}

Block-structured AMR codes define the solution across a hierarchical representation of different levels of resolution.  At each level, the solution is defined over a region represented by the
union of a collection of non-overlapping rectilinear regions with associated data containers.  These data containers can hold mesh data, particles, information describing an embedded boundary geometry or other data used to represent the solution. In this section, we introduce the data structures that describe the grid hierarchy and how the hierarchy is generated.

\subsection{Basic data structures}

\commentout{AMReX uses a collection of distributed boxes that are distributed across MPI ranks.}
AMReX uses a number of data abstractions to represent the grids at a given level.
The basic objects are:
\begin{enumerate}
    \item \texttt{IntVect}: a dimension-sized list of integers representing a spatial location in index space.  The use of \texttt{IntVect} allows the basic grid description to be dimension-independent.
    \item \texttt{Box}: a logically rectangular (or rectangular cuboid, in 3D) region of cells defined by its upper and lower corner indices. A \texttt{Box} also has an \texttt{IndexType}, a specialized \texttt{IntVect} that describes whether the indices refer to cell centers, faces, edges or nodes.
    \item \texttt{BoxArray}: a collection of \texttt{Box}es non-intersecting
    \commentout{that describes a complete mesh}
    at the same resolution. Each individual \texttt{Box} in a \texttt{BoxArray} can be stored, accessed, copied and moved independently of the other boxes.
    \item \texttt{DistributionMapping}: a vector that lists the MPI rank that the data associated with each \texttt{Box} in a \texttt{BoxArray} will be defined on. 
    \commentout{A \texttt{BoxArray} and \texttt{DistributionMapping} describe the memory layout of any grid in AMReX.}
\end{enumerate}

In an AMReX application, the \texttt{BoxArray} and \texttt{DistributionMapping} at each level describe how the data at that level is decomposed and where it is allocated.  The relationship between boxes at different levels, thus resolution different levels, is defined by a refinement ratio.  In a single application, the refinement ratio between any two levels may differ but it must be prescribed at runtime. 
\commentout{This data species the geometric relationship between boxes at different levels that is needed for applications.}
To facilitate efficient communication across MPI ranks, the \texttt{BoxArray} and \texttt{DistributionMapping} at each level, also referred to as the metadata, are stored on every MPI rank.

\commentout{In addition to the hierarchy of grid description, }
AMReX also includes distinct objects commonly used to describe other critical simulation parameters.  Of particular importance are:
\begin{itemize}
%    \item \texttt{Array4}: a GPU friendly indexing object, that allows users to access data in a readable, Fortran-like (i, j, k) format.
    \item \texttt{Geometry}: stores and calculates parameters that describe the relationship between index space and physical space.  The \texttt{Geometry} object holds information that is the same at all levels, such as periodicity and coordinate system, but also level-specific information such as mesh size.  A multi-level AMReX-based simulation will typically have a \texttt{Geometry} object at each level.
    \item \texttt{BCRec}: a multi-dimensional integer array that defines the physical boundary conditions for the domain.  A multi-level AMReX-based simulation will only need one \texttt{BCRec}.
%    \item \texttt{FluxRegister}: stores and adjusts fluxes at coarse-fine interfaces
%    \item \texttt{StateData}: a component of \texttt{Amr} that stores a pair of \texttt{MultiFabs}, one for the previous time step and one for the current time step.
%    \item \texttt{ParticleContainer}: container of local \texttt{Particle}s, optimized for both speed and memory.
\end{itemize}

%Because of the structured nature of block-structured algorithms, this data gives AMReX the capability to provide a wide array of portable and efficient AMR calculations.  For example, AMReX functions are available to fill ghost cells, implement boundary conditions, perform reduction operations and interpolate in time or from coarse levels to fine levels.  By buildin

%\MarginPar{Is there a list of functions we'd like to add to this? Maybe point out specific fucntions? Which ones?}

\subsection{Regridding and load balancing}

AMR algorithms dynamically change the resolution of the mesh to reflect the changing requirements of the problem being solved.  
AMReX provides support for the creation of new grids or the modification of existing grids (regridding) at level $\ell+1$ and above at user-specified intervals. To do so, 
\commentout{Given data at level $\ell-1$}
an error estimate is used to tag individual cells above a given tolerance at level $\ell$, where the error is defined by user-specified routines.  Typical error criteria might include,
first or second derivatives of the state variables or some type of application-specific heuristic.

The tagged cells are grouped into rectangular grids at level $\ell$ using the clustering algorithm given in \cite{bergerRigoutsos:1991}.
These rectangular patches are refined to form the grids at level $\ell+1$.
Large patches are broken into smaller patches for distribution to multiple 
processors based on a user-specified \texttt{max\_grid\_size} parameter; AMReX enforces that no grid is longer in any direction than \texttt{max\_grid\_size} cells.
AMReX also provides the option to enforce that the dimensions of each newly created grid is divisible by a specific factor, \texttt{blocking\_factor}; this is used to improve multigrid efficiency for single-level solves. (In this case the grid creation algorithm is slightly different; we refer the interested reader to the AMReX documentation: \cite{AMREX:2019,AMREX:Docs}.)
The refinement process is repeated until either an error tolerance criterion is satisfied or a specified maximum level is reached.  
Finally, we impose the proper nesting requirements that enforce that level $\ell+1$ is strictly contained in level $\ell$ except at physical boundaries.
A similar procedure is used to generate the initial grid hierarchy.
Once created, the \texttt{BoxArray} of new grids at each level is broadcast to every MPI rank.

\commentout{Once the new grids have been created, the new grid hierarchy needs to be distributed to the MPI ranks.}
%Load-balancing is so critical to AMR codes that even initial data distributions implement a default strategy.

When creating a \texttt{DistributionMapping} associated with a \texttt{BoxArray} at a level it is important to equidistribute the work to the nodes. Some type of cost function that is assumed or prescribed by the user is used to estimate the work.  The default methodology assumes a cost function proportional to the number of cells per grid, and implements a Morton-ordering space filling curve (SFC) algorithm to define an initial data distribution across the MPI ranks.
Once data has been allocated, applications can calculate an improved \texttt{DistributionMapping} throughout runtime based on any desired load-balancing parameter, such as data size, an estimation of the work, a user-defined cost function or a timer.

Prescribed cost functions are often less than optimal.  For GPUs, both CUPTI timers and efficient hand-written timers have been developed to accurately capture device kernel times respective to the device.  This eliminates launch delays and other latencies from the resulting timer that prevent accurate distribution.
  
\begin{figure}[h]
    \centering
    \includegraphics[width=0.48\textwidth]{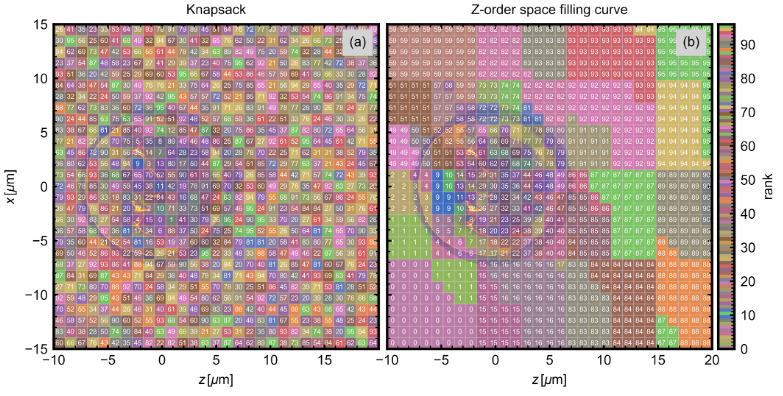}
    \caption{Comparison of the data distribution from the knapsack load balancing algorithm (left) and SFC (right).  Image courtesy of Michael Rowan.}
    \label{fig:load}
\end{figure}

AMReX provides a knapsack algorithm in addition to a space-filling curve algorithm for load balancing simulations across MPI ranks.  The knapsack algorithm creates the most optimal distribution possible with respect to the load-balancing parameter, while the SFC tries to build contiguous spatial regions with near uniform costs.  See Figure \ref{fig:load} for an illustration of how the two approaches distribute data.
\commentout{Once a new distribution mapping has been calculated, a simple call, the new distributed data containers can be filled from the old hierarchy.}
%\MarginPar{not sure i follow the last bit here.  tradeoff is buffering versus frequency.  also say something about how to fill in data}
%\MarginPar{Have a recent picture from Micheal Rowan's upcoming SC paper showing the difference between SFC and knapsack, if desired. It's uploaded as "LoadBalancing.png".}
%to \texttt{Redistribute} implements the calculated data mapping.  Load balancing can be expensive for large simulations, so characteristics like how often a new distribution is generated and the amount of improvement needed to trigger a redistribution must be tuned to provide optimal performance.

\section{Mesh data and iterators}

At the core of the AMReX software is a flexible set of data structures that can be used to represent block-structured mesh data in a distributed memory environment. Multilevel data on the AMR grid hierarchy are stored in a vector of
single level data structures.  
\commentout{Note that the grids on a single level
consist of multiple rectangular subdomains.}
In this section we describe the data containers for
mesh data.  We also discuss operations supported on these data structures including iterators for operations at a level, a communications layer to handle ghost cell exchange and data distribution and tools for operations between levels.
%\MarginPar{3 pages}

\subsection{Containers for mesh data}

The basic single level
data structure in AMReX is a C\texttt{++} class
template, \texttt{FabArray<FAB>}, a distributed data structure for a collection of \texttt{FAB}s,
one for each grid.  The template parameter \texttt{FAB} is
usually a multidimensional array class such as \texttt{FArrayBox} for
floating point numbers.  Most AMReX applications use a specialization
of \texttt{FabArray}, called a \texttt{MultiFab}, that is used for
storing floating point data on single level grids. 
However, \texttt{FabArray} can also be used to store integers, complex
numbers or any user defined data structures.

An \texttt{FArrayBox} contains the \texttt{Box} that defines its valid region,
and a pointer to a multidimensional array.
For three-dimensional calculations, the data accessed by the pointer in an \texttt{FArrayBox} or its base
class \texttt{BaseFab<T>} is a four-dimensional array.  The first three dimensions correspond to the spatial dimensions; the fourth dimension is for components. An important feature
of an \texttt{FArrayBox} is that extra space is allocated to provide space for ghost cell 
data, which is often required for efficient stencil operations. Ghost cell data are typically filled by interpolating from coarser data, copying from other \texttt{FArrayBox}es at the same level, or by imposing boundary conditions outside the domain. 
\commentout{with data from the grid hierarchy to facilitate the types of stencil operations often used in PDE simulations.} 
\commentout{AMReX includes a simple query mechanism to determine the valid region and what portion corresponds to ghost cells.}
The data in an \texttt{FArrayBox} (including ghost cells) are stored in a contiguous chunk of memory
with the layout of the so-called struct of arrays (SoA) (i.e., struct of
3D arrays, one for each component).  
\commentout{Being contiguous in memory makes it interoperable with C and Fortran, and moreover easy for GPU computing.}

\commentout{To facilitate the use of GPUs,}
AMReX provides a class template, \texttt{Array4<T>}, that can be used to access the
data in an \texttt{FArrayBox} or other similar objects.  
This class does not own the data and is not responsible for allocating or freeing memory; it simply provides a reference to the data. This property makes it suitable for being captured by a C++ lambda function without memory management concerns.
\commentout{and fits nicely with our GPU strategy based on C++ lambda expressions.}  
The \texttt{Array4} class has
an \texttt{operator()} that allows the user to access it with Fortran
multi-dimensional array like syntax.  The SoA layout
of \texttt{FArrayBox} is typically a good choice for most applications
because it provides unit stride memory access for common
operations.  Nevertheless, some algorithms (e.g., discontinuous
Galerkin methods) perform computations that involve relatively large
matrices within a single element, and an array of struct (AoS) is a
better layout for performance.  This can be achieved
with \texttt{BaseFab<T>}, where \texttt{T} is a struct.

\subsection{Iterators}

%\MarginPar{intro what this is about.  owner computes, etc.}
AMReX employs an owner-computes rule to operate on single AMR level data structures such as \texttt{MultiFab}s and \texttt{FabArray}s.  A data iterator, \texttt{MFIter}, is provided to loop over the single box data structures
such as \texttt{FArrayBox}s and \texttt{BaseFab}s in a \texttt{MultiFab} / \texttt{FabArray} with the option of logical tiling.  The iterator can be used to access data from
multiple \texttt{MultiFab}s / \texttt{FabArray}s built on the same \texttt{BoxArray} with
the same \texttt{DistributionMapping}.

For CPU computing, tiling provides the benefit of
cache blocking and parallelism for OpenMP threading.  The iterator
performs a loop transformation, thereby improving data locality and relieving
the user from the burden of manually tiling \citep{BoxLibTiling}.
For GPU computing, tiling is
turned off by default because exposing more fine-grained parallelism
at the cell level and reducing kernel launch overhead are more
important than cache blocking.

\subsection{Operations on a level}

Many AMReX-based applications use stencil operations that require access to the data on neighboring and/or nearby cells.  When that data lies outside the valid region of the patch being operated on, pre-filling ghost cells is an efficient way to optimize communication.  The \texttt{Box}es stored in \texttt{FabArray}s determine the ``valid" region.  The multidimensional array associated with each \texttt{FArrayBox} covers a region larger than the ``valid" region by \texttt{nGrow} cells on the low and high sides in each direction, where \texttt{nGrow} is specified when the \texttt{MultiFab} is created.  
AMReX provides basic communication functions for ghost cell exchanges of data in
the same \texttt{FabArray}.
It also provides routines for copying data between two
different \texttt{FabArray}s at the same level.
These routines use the metadata discussed in the previous section to compute
where the data is located.  Data between different MPI ranks are organized into buffers
to reduce the number of messages that need to be sent.  Techniques to optimize the
communication are discussed in mote detail below.

%\Ann{A little more detail above about how this has been made fast would be good here since this is such a common operation}

\subsection{Operations between levels}

AMReX also provides functions for common communication patterns associated with
operations between adjacent levels.  The three most common types of inter-level
operations are interpolation from coarse data to fine, 
restriction of fine data to coarse, and explicit refluxing 
at coarse/fine boundaries. \commentout{, and synchronization of implicit
discretizations at coarse/fine boundaries.}  

Interpolation is needed to fill fine level ghost cells that are inside the domain but can't be filled by copying from another patch at the same level, and is also needed in regridding to fill regions at the fine level that weren't previously contained in fine
level patches.  Note that to avoid unnecessary communication we only obtain
coarse data to interpolate to the fine level when fine data is not available.
%Note that to avoid unnecessary communication needed coarse data are
%communicated and then interpolated to the fine level , only when the fine data
%cannot be obtained from the fine level,.
Restriction is typically used to synchronize data between
levels, replacing data at a coarser level with an average or injection of
fine data where possible.   A number of interpolation and restriction operations are
provided within AMReX for cell-centered, face-centered or nodal data; a developer can 
also create their own application-specific interpolation or restriction operation.
For these operations, necessary communication is done first and then interpolation / restriction
is done locally, making it straightforward for users to provide their own functions.
AMReX only needs to know the width of the stencil to gather the necessary amount of data.
%\Weiqun{Because inter-level interpolation involves a sequence of data
%communciation and local interpolation operations, AMReX provides
%functions that can take user provided interpolation functions.}

Refluxing effectively synchronizes explicit fluxes across coarse/fine boundaries, and is relevant only for cell-centered data being updated with face-centered fluxes.
Specialized data structures, referred to as \texttt{FluxRegister}s, 
hold the differences between fluxes on a coarse level and time- and space-averaged fine level fluxes.
Once the coarse and fine level have reached the same simulation time, the data in the \texttt{FluxRegister}s are used to update the data in the coarse cells immediately next to the coarse-fine boundaries but not covered by fine cells.  This operation is common in multi-level algorithms with explicit flux updates.  \texttt{FluxRegister}s are also used to hold data needed to synchronize implicit discretizations of parabolic and elliptic equations. 

%\Ann{I commented out the discussion of FluxRegisters for implicit stuff -- as is, it's too confusing, and I'm not sure it warrants the space to fully explain.}
\commentout{For linear solvers used in implicit and semi-implicit discretizations of 
parabolic and elliptic equations, synchronization at coarse/fine
boundaries is needed to maintain consistent properties like divergence
constraint.  \texttt{FluxRegister}s are also used to store the data needed to form the
right hand sided for implicit synchronization associated with implicit discretizations.}

\subsection{Communication}

%\MarginPar{need to obtain data from other nodes.  caching}

AMReX grids can have a complicated layout, 
which makes the communication meta-data
needed 
for ghost cell exchange and inter-level communication non-trivial
to construct.  AMReX uses a hash-based algorithm to perform
intersections between \texttt{Box}es. The hash is constructed in an
$O(N)$ operation, where $N$ is the number of global
\texttt{Box}es, and cached for later use.  It then takes only
$O(n)$ operations to construct the list of source and destination
\texttt{Box}es for communication, where $n$ is the number of local \texttt{Box}es.
Furthermore, the communication meta-data is also cached for reuse.

AMReX supports GPU-aware MPI on GPU machines, if available.  To reduce
latency, the data needed to be communicated through MPI are aggregated
into communication buffers.  For packing the buffer, we need to copy
data from slices of multi-dimensional arrays to the one-dimensional
buffer, whereas for unpacking the buffer, data are copied from the
one-dimensional buffer to slices of multi-dimensional array.  The
straightforward approach of the copying is to launch a GPU kernel for
each slice.  Unfortunately this simple approach is very expensive
because there are often hundreds of very small GPU kernels.  In
AMReX, we have implemented a fusing mechanism that merges all these
small GPU kernels into one kernel, thus significantly reducing kernel
launch overhead.

\section{Performance portability}
Performance portability across different architectures has been identified as a major concern for the ECP. Most AMReX-based application codes either use or have plans to use a variety of platforms, including individual workstations, many-core HPC platforms such as NERSC's Cori, and accelerator-based supercomputers such as Summit and Frontier at OLCF and Aurora at ALCF. In this section we discuss various features that enable applications to obtain high performance on different architectures without substantial recoding.

\subsection{On-node parallelism}

AMReX is based on a hierarchical parallelism model. 
At a coarse-grained level,
the basic AMReX paradigm is based on distribution of one or more patches of data to each node with an owner-computes rule to allocate tasks between nodes. For many use cases, a node is divided into a small number of MPI ranks and the coarse-grained distribution is over MPI ranks.  For example, on a system with 6 GPUs per node, the node would typically have 6 MPI ranks.
The nodes on modern architectures all have hardware for parallel execution within the node but there is considerable variability in the the details of intranode parallel hardware.  For code executing on CPUs,  AMReX supports logical tiling for cache re-use using OpenMP threading.  Tile size can be adjusted at run time to improve cache performance; tile size can also vary between operations.  AMReX includes both a standard synchronous strategy for scheduling tiles as well as an asynchronous scheduling methodology. AMReX also provides extensive support for kernel launching on GPU accelerators (using C++ lambda functions) and for the effective use of managed memory, that allows users to control where their data is stored.  While much of the internal AMReX functionality currently uses CUDA/HIP/DPC++ for maximum performance on current machines, AMReX supports the use of CUDA, HIP, DPC++, OpenMP or OpenACC in AMReX-based applications. Specific architecture-dependent aspects of the software for GPUs are highly localized, enabling AMReX to easily support other GPU architectures.
%\MarginPar{strawman}

To isolate application code from the nuances of a particular architecture, AMReX provides an abstraction layer consisting of a number of \texttt{ParallelFor} looping constructs, similar to those provided by  Kokkos \citep{kokkos} or RAJA \citep{raja} but tailored to the needs of block-structured AMR applications. Users provide a loop body as a C++ lambda function that defines the task to be performed over a set of cells or
particles.  The lambda function is then used in a kernel with a launching mechanism provided by CUDA, HIP or DPC++ for GPU or standard C++ for CPU.  The details of how to loop over the set of objects, and in particular how to map iterations of the loop to hardware resources such as GPU threads, are hidden from users. By separating these details from the loop body, the same user code can run on multiple platforms. In addition to 1D loops, users can also specify 3D and 4D loops using by passing in an \texttt{amrex::Box} as the loop bounds.  

Internally, the \texttt{ParallelFor} construct in AMReX is implemented using CUDA, HIP, and DPC++ as the parallel backend. A key difference between AMReX's \texttt{ParallelFor} functions and those provided by Kokkos or RAJA is the way OpenMP is handled. When using multi-threading, the AMReX \texttt{ParallelFor} does not include any OpenMP directives (it does, however, incorporate vectorization through \#pragma simd). Rather, OpenMP is incorporated at the MFIter level, as shown in Listing~\ref{lst:mfiter_parfor}. When this code is compiled to execute on CPUs with OpenMP, the \texttt{MFIter} loop includes tiling and the \texttt{ParallelFor} translates to a serial loop over the cells in a tile. However, when compiling for GPU platforms, tiling at the \texttt{MFIter} level is switched off, and the \texttt{ParallelFor} translates to a GPU kernel launch. 

\begin{lstlisting}[language=C++,frame=single,basicstyle=\footnotesize,label={lst:mfiter_parfor},caption={Example of ParallelFor.  This code can be compiled to run on CPU with OpenMP or GPU with CUDA, HIP, or DPC++.}]
#pragma omp parallel if (Gpu::notInLaunchRegion())
for (MFIter mfi(mf,true); mfi.isValid(); ++mfi) {
  Array4<Real> const& fab = mf.array(mfi);
  ParallelFor(mfi.tilebox(),
    [=] AMREX_GPU_DEVICE (int i, int j, int k) {
      fab(i,j,k) *= 3.0;
  });
}
\end{lstlisting}

% fused kernel launches -- why and how

\subsection{Parallel Reductions}
Parallel reductions are commonly needed in scientific computing. For example, in an explicit hydrodynamics solver, one needs to calculate the time step by examining the CFL constraint in each cell and then finding the minimum over all cells. 
Similarly, in linear solvers, one needs to measure norms of the global residual.
\commentout{Or, in a molecular dynamics code, one may want to compute diagnostics such as the total number of collisions, or the closest distance between two particles.} Performing reductions efficiently in parallel in a way that is portable between CPU and various GPU platforms is non-trivial. To aid in this task, AMReX provides generic functions for performing reduction operations in a performance-portable way. These functions can be used at the level of contiguous arrays by passing in data pointers, or they can work on higher-level AMReX data containers to e.g. perform reductions over all the cells in a \texttt{MultiFab}, or all the particles in a \texttt{ParticleContainer} (see \nameref{sec:particles} for more information about particles in AMReX).

A feature of the AMReX implementation of parallel reduce is that we provide an API for performing multiple reductions in one pass using a \texttt{ReduceTuple} datatype. See Listing \ref{lst:reduction} for a code example. The code in this snippet computes the Sum, Min, and Max of all the \texttt{Real} data in a \texttt{MultiFab}, as well as the Sum of all the \texttt{int} data in an \texttt{iMultiFab} (\texttt{Long} is used in the reduction to avoid overflow). Any arbitrary combination of data types and reduction operators can be applied in this manner. When running on GPUs, all these operations would be done in a single kernel launch. 

\begin{lstlisting}[language=C++,frame=single,basicstyle=\footnotesize,label={lst:reduction},caption={Performing parallel reduce on mixed data types.}]
ReduceOps<ReduceOpSum, ReduceOpMin, ReduceOpMax, ReduceOpSum> reduce_op;
ReduceData<Real, Real, Real, Long> reduce_data(reduce_op);
using ReduceTuple = typename decltype(reduce_data)::Type;

for (MFIter mfi(mf); mfi.isValid(); ++mfi)
{
    const Box& bx = mfi.validbox();
    auto const& fab = mf.array(mfi);
    auto const& ifab = imf.array(mfi);
    reduce_op.eval(bx, reduce_data,
    [=] AMREX_GPU_DEVICE (int i, int j, int k) -> ReduceTuple
	{
        Real x =  fab(i,j,k);
	      Long ix = static_cast<Long>(ifab(i,j,k));
        return {x,x,x,ix};
    });
}
\end{lstlisting}

Note that the code in the above snippet does not perform MPI-level reductions. If that is needed, for convenience AMReX provides wrappers such as \texttt{ParallelDescriptor::ReduceRealMin} and \texttt{ParallelDescriptor::ReduceLongSum}, so that the same code can work whether or not it is compiled with MPI support.

\subsection{Memory management}
Allocating and deallocating memory can be extremely costly, especially on CPU-GPU platforms where data needs to be copied back and forth between multiple memory spaces.  Furthermore, different architectures support different types of memory models.  AMReX includes a variety of memory pools, referred to as \texttt{Arena}s, to improve the performance of memory activities and properly track and handle memory allocations.  \texttt{Arena}s allocate a large block of memory during initialization and provide data pointers to pieces of that pre-allocated space as needed.  This minimizes expensive allocation calls, provides a portability wrapper around memory operations and allows tracking of memory use across a simulation.  In most cases, \texttt{Arena}s are used automatically according to AMReX's memory strategy.  For example, when creating a new \texttt{MultiFab}, the mesh data is allocated with the default AMReX \texttt{Arena}, \texttt{The\_Arena}, unless otherwise specified by the user.
%\Kevin{change to: changed by the user?}

Based on the system architecture, AMReX predefines an \texttt{Arena} for each available memory type during initialization.  A variety of \texttt{Arena} types are provided, including slab allocation, buddy memory and best-fit, and they can be targeted to any desired memory type, which currently includes CPU memory, pinned memory, GPU device memory and GPU managed memory.  Users can build their own \texttt{Arena}s, either to track a particular subset of memory usage or to manage memory according with a user-specific strategy.  AMReX also provides specialized \texttt{std::vector} variations that track particle memory use as well as implement an improved memory allocation strategy.

AMReX's default memory management strategy for CPU-GPU systems is based on placing floating point data on the device and moving it as little as possible.  Using this paradigm, metadata is usually allocated to the CPU to facilitate efficient host side solution control and pinned memory is used for temporary data that will be transferred between host and device.  This memory management strategy is implemented automatically, and users only interact directly with \texttt{Arena}s when it becomes necessary to deviate from this strategy.  AMReX \texttt{Arena}s include a robust design API so manually controlling deviations is simple.  For example, by default mesh and particle data are allocated in managed memory, due to its convenience to developers and negligible overhead.  But, this default can be overridden for any specific \texttt{MultiFab}, \texttt{iMultiFab} or \texttt{ParticleContainer} object during the object's constructions.  The default memory pool can also be changed to traditional device memory through a runtime flag.  Additional runtime options include setting the initial allocation sizes of the \texttt{Arena}s and aborting if device memory is oversubscribed, allowing users to adjust the AMReX memory management paradigm for their unique needs.

\subsection{Example}
%xxx Not sure where to put this.  I put this in a separate file for now xxx

As an illustration of the performance of AMReX on a simple example, we carried out a weak scaling study on Summit (https://www.olcf.ornl.gov/summit/) using an explicit
hydrodynamics code for solving the compressible Euler equations.  The time-stepping strategy uses a second-order Runge-Kutta method on each level with subcycling in time between levels.
A piecewise linear reconstruction method with a monotonized central slope limiter
is used to extrapolate the primitive variables to cell faces, and an approximate
Riemann solver based on the two-shock approximation is employed to
compute fluxes.  We have performed a series of tests simulating the
Rayleigh-Taylor instability.  The runs have a base level and two
refined levels, each with a refinement ratio of 2.  On each node, there are
$128 \times 128 \times 256$ cells on level 0.  For multi-node runs,
the domain is replicated in the $x$ and $y$-directions.  In the initial
model, levels 1 and 2 occupy 12.5\% and 6.25\% of the domain,
respectively.  Regridding is performed every 2 steps on each fine
level.  

\begin{figure}[ht]
    \centering
    \includegraphics[width=0.45\textwidth]{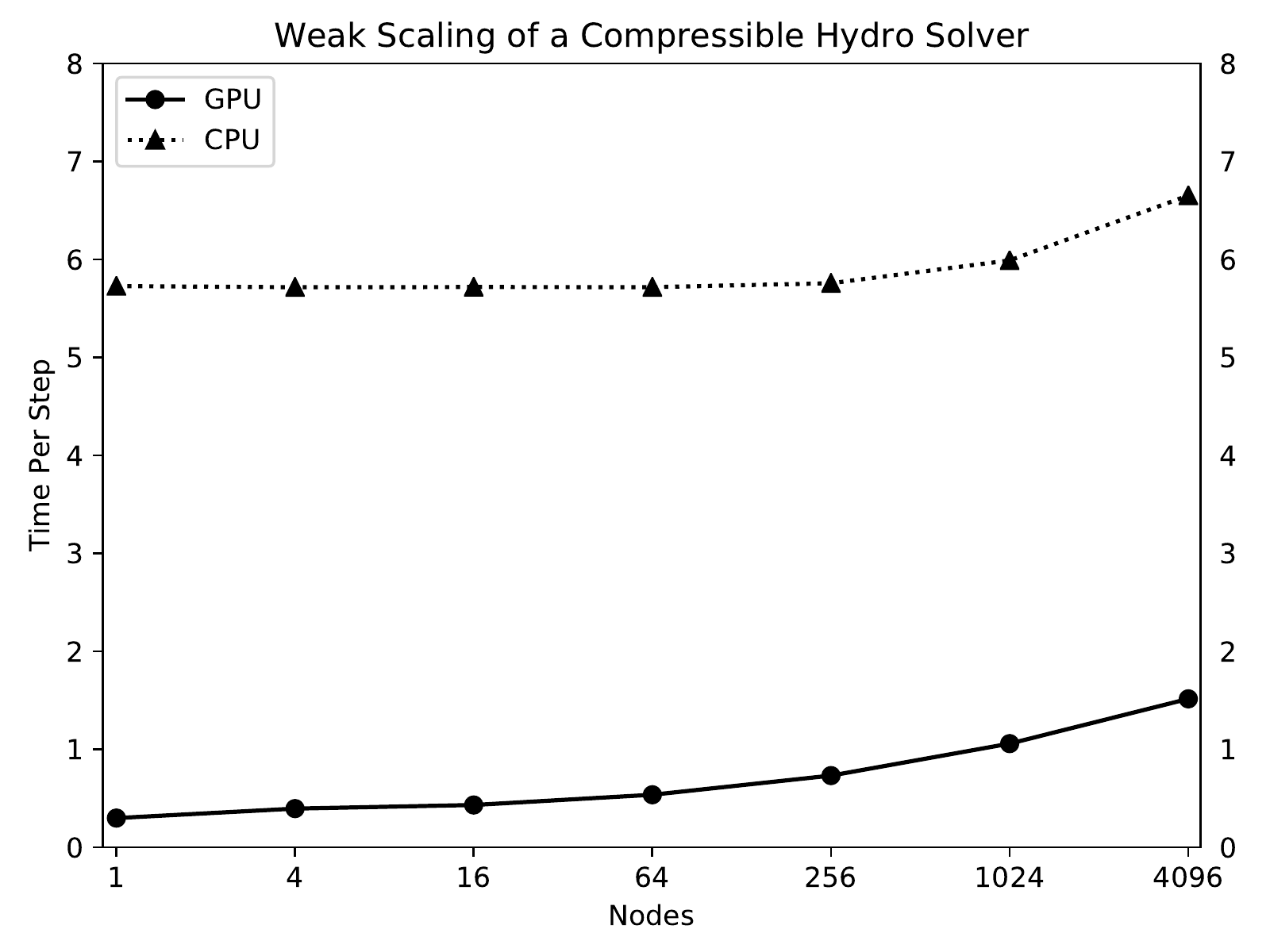}
    \caption{Weak scaling study of a compressible hydrodynamics
      solver.  Time per step is shown for a series of GPU and CPU
      runs.  The AMR runs have three total levels.}
    \label{fig:cns}
\end{figure}

Figure~\ref{fig:cns}
shows the results of a series of runs on Summit.  We define the speedup as the time per step using CPUs only divided by the time per step using 6 GPUs per node. 
For single node runs, there is a $\sim 60 \times$ speedup for the computational kernels themselves, and a 19$\times$ overall speedup with GPUs.  For the
runs using 4096 nodes (24576 GPUs), a 4.4$\times$ speedup is obtained.  
The reduced speedup is due
to increased communication costs relative to compute costs when using GPU accelerators; the effect is so large here because relatively little floating-point work needed to advance the solution in each cell.

\section{Particles}
\label{sec:particles}
In addition to mesh data, many AMReX applications use particles to model at least some portion of the physical system under consideration. At their simplest, particles can be used as passive tracers that follow a flow field, tracking changes to properties such as chemical composition but not feeding back into the mesh-based solution. In most cases, however, the particles in AMReX applications actively influence the evolution of the mesh variables, through, for example, drag or gravitational forces. Finally, particles can directly influence each other without an intervening mesh, for example through particle-particle collisions or DSMC-style random interactions. AMReX provides data containers, iterators, parallel communication routines, and other tools for implementing the above operations on particle data.

Particles present a different set of challenges than structured grid data for several reasons. First, particle data is inherently dynamic - even between AMR regridding operations, particles are constantly changing their positions. Additionally, the connectivity of particle data (e.g. nearest neighbors) is not as simple and in general cannot be inferred from metadata alone - rather, the particle data itself must be examined and processed. Finally, data access patterns are more irregular, in that (unless special sorting is employed) particles that are next to each other in memory do not necessarily interact with the same set of grid cells or with the same set of other particles. In what follows, we describe the particle data structures and supporting operations provided by AMReX, and how they help users manage these challenges.   

\subsection{Data structures and iterators}

AMReX provides a \texttt{ParticleContainer} class for storing particle data in a distributed fashion. Particles in AMReX are associated with a block-structured mesh refinement hierarchy based on their position coordinates. Internally, this hierarchy is represented by a \texttt{Vector<BoxArray>} that describes the grid patches on each level, as well as a \texttt{Vector<DistributionMapping>} that describes how those grids are distributed onto MPI ranks. Particles are then assigned to levels, grids and MPI ranks spatially, by binning them on a given level using the appropriate cell spacing and physical domain offset. Storing the particles in this fashion allows them to be accessed and iterated over level-by-level, which is convenient for many AMR time-stepping approaches. Additionally, splitting particles onto grids and then assigning multiple grids per MPI rank allows load balancing to performed by exchanging grids among the processes until an even number of particles is achieved. Note also that although particles in AMReX always live on a set of AMR grids, this does not need to the same set of grids that is used to store the mesh data; in principle, the particle grids can be different and can be load balanced separately. See \nameref{sec:particles:dual_grid} for more information.

Once the particles are stored in the \texttt{ParticleContainer}, they can be iterated over in parallel in a similar manner to the \texttt{MFIter} for mesh data. See Listing \ref{lst:pariter} for a code sample. Specifically, when an MPI rank executes a \texttt{ParIter} loop, it executes the loop body once for each grid (or tile, if tiling is enabled, see \nameref{sec:particles:data_layout}) on the level in question. By launching the same compute kernel on each grid, all the particles on the level can be updated in a manner that separates the numerical operations from the details of the domain decomposition. Inside the \texttt{ParIter}, the same \texttt{ParallelFor} constructs can be used to operate on particles in a portable way. The 1D version of \texttt{ParallelFor}, which takes a number of items to iterate over instead of a \texttt{Box}, is particularly useful for particle data. 

\begin{lstlisting}[language=C++,frame=single,basicstyle=\footnotesize,label={lst:pariter},caption={Example of ParIter loop}]
#pragma omp parallel if (Gpu::notInLaunchRegion())
for(ParIter pti(pc, lev, TilingIfNotGPU()); 
    pti.isValid(); ++pti)
{
    auto& tile = pti.GetParticleTile();
    const auto np = tile.numParticles();
    auto pdata = tile.getParticleTileData();

    amrex::ParallelFor( np, 
    [=] AMREX_GPU_DEVICE (int i)
    {
        some_function(pdata, i);
    });
}
\end{lstlisting}

\subsection{Data Layout}
\label{sec:particles:data_layout}
AMReX supports particle types that are arbitrary mixtures of real (i.e., floating point data of either single or double precision) and integer components. For example, a particle used to represent a star in a 3D N-body calculation might have four real components, a mass, and three velocity components, while a particle that a represents an ion in a fuel cell might have an integer component that tracks its ionization level. In general, the particles in a given container must all have the same type, but there is no limit to the number of different types of \texttt{ParticleContainer}s that can be in the same simulation. For example, a single calculation might have one type of particle that represents collisionless dark matter, another that represents neutrinos, and another that represents active galactic nuclei, each of which would model different physics and exhibit different behavior. 

There is also the question of how the various components for the particles on a single grid should be laid out in memory. In an Array-of-Structs (AoS) representation, all the components for particle 1 are next to each other, followed by all the components for particle 2, and so on. In a Struct-of-Arrays (SoA) representation, all the data over all the particles for component 1 would be next to other, followed by all the data for component 2, and so on. AMReX allows users to specify a mixture of AoS and SoA data when defining the particle type. For example, consider an application where, in addition to position and velocity, each particle stores the mass fraction of a large number of chemical components. For many operations, such as advancing the particle positions in time or assigning particles to the proper grid in the AMR hierarchy, the values of the chemical concentration are irrelevant. If all of these components were stored in the particle struct, the data for the position and velocity of all the particles would be spread out in memory, resulting in an inefficient use of cache for these common operations. By separating out these components from the core particle struct, we can avoid reading them into cache for operations in which they are not needed. 

The AMReX particle containers also support a tiling option. When tiling is enabled, instead of each grid having its own separate set of particle data structures, the grids will be further subdivided into tiles with a runtime-specified size. Unlike with mesh data, with particle data the tiles are not just logical, they actually change the way the data is laid out in memory, with each tile getting its own set of set of \texttt{std::vector}-like data structures. This tiling strategy can make working sets smaller and more cache-friendly, as well as enable the use of OpenMP for particle-mesh deposition operations, in which race conditions need to be considered (see \nameref{sec:particles:particle_mesh} for more information).

\subsection{Particle Communication}

There are two main parallel communication patterns that arise when implementing particle methods with AMReX. The first is redistribution. Suppose that at a given time in a simulation, the particles are all associated with the correct level, grid, and MPI rank. The particle positions are then updated according to some time-stepping procedure. At their new positions, the particles may no longer be assigned to the correct location in the AMR hierarchy. The \texttt{Redistribute()} method in AMReX takes care of assigning the correct level, grid, and tile to each particle and moving it to the correct place in the \texttt{ParticleContainer}, including any necessary MPI communication. We provide two forms of \texttt{Redistribute()}, one in which the particles have only moved a finite distance from the ``correct" grids and thus are only exchanged between neighboring MPI ranks, and another where the particles can in principle move between any two ranks in the MPI communicator. The former is the version most often used during time evolution, while the latter is mostly useful after initialization, regridding, or performing a load-balancing operation. 

The process of particle/grid assignment is done as follows. First, consider a single level. Assigning a particle to a grid reduces to 1) binning the particle using the level's cell spacing, and 2) identifying the box in the \texttt{BoxArray} that contains the associated cell. Internally, this operation is implemented using a uniform binning strategy to avoid a direct search over all the boxes on the level. By binning boxes using the maximum box size as the bin size, only neighboring bins need to be searched to find the possible intersections for a given cell. Additionally, this operation can be performed purely on the GPU by using a flattened map data structure to store the binned \texttt{BoxArray}. 

To extend this operation to multiple levels, we consider each level in a loop, from finest to coarsest. Usually, we associate the particles with the finest level possible. The exception is when subcycling in time is involved. In this case, we provide the option to exclude certain levels from the \texttt{Redistribute()} operation, and to leave the particles alone if they are less than a specified number of cells outside the coarse/fine boundary.

The second communication pattern that comes up for particle data is the halo pattern, similar to filling ghost cells for mesh data. In this operation, each grid or tile obtains copies of particles that are within some number of cells outside its valid region. This is commonly needed in applications with direct particle-particle collisions, since there is no intermediate grid that can be used to communicate accumulated values (in contrast to the pattern described in \nameref{sec:particles:particle_mesh}). The AMReX terminology for these copies of particles from other grids is \texttt{NeighborParticles}. As with \texttt{Redistribute()}, any necessary parallel communication is handled automatically, and when AMReX is compiled with GPU support the operation runs entirely on the device.  

When using neighbor particles, it is common to compute a halo of particles (i.e., identify which particles need to be copied to which other grids) once, and then reuse it for several time steps before computing a new halo. To enable this, AMReX separates computing the halo and filling it with data for the first time (the \texttt{FillNeighbors} operation) from reusing the existing halo but replacing the data with the most recent values from the ``valid" particles on other grids (the \texttt{UpdateNeighbors} operation). Additionally, AMReX provides a \texttt{SumNeighbors} function that takes data from the ghosted particles and adds the contribution back to the corresponding  ``valid" particles. 

\begin{figure}[h]
    \centering
    \includegraphics[width=0.45\textwidth]{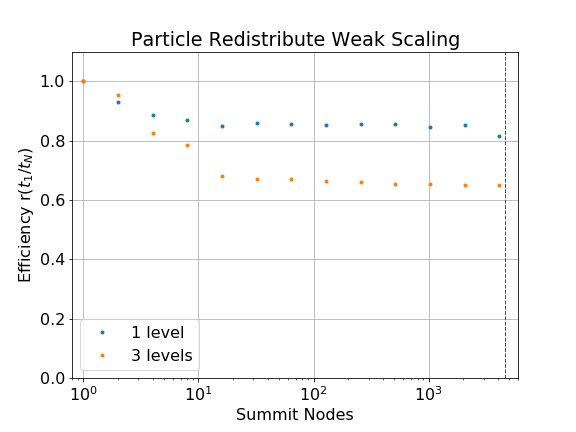}
    \caption{Weak scaling study of an AMReX particle redistribution benchmark on Summit. The $x$-axis is the number of nodes, while the $y$-axis shows parallel efficiency, i.e. the run time on 1 node divided by the run time on a scaled up version of the benchmark on $N$ nodes. The total number of nodes on Summit is indiciated by the red vertical line. All runs used 6 MPI ranks per node with 1 GPU per MPI rank. Results are shown for both a 1-level and a 3-level version of the benchmark.}
    \label{fig:redistribute_scaling}
\end{figure}

The particle communication routines in AMReX been optimized for Summit and can weak scale up to nearly the full machine. Figure \ref{fig:redistribute_scaling} shows the results of a particle redistribution scaling study on up to 4096 Summit nodes, with and without mesh refinement. In this test, particles are initialized uniformly throughout the domain and then moved in a random direction for 500 time steps, calling \texttt{Redistribute()} after each one. The initial drop in weak scaling efficiency from 1 to $\sim 16$ nodes is due to the fact that in 3D, the average number of communication partners per MPI rank grows until you have $\approx 27$ neighbors per grid. After this point, the scaling is basically flat up to 4096 nodes (24576 MPI tasks). Finally, although this benchmark tests \texttt{Redistribute()} rather than \texttt{FillNeighbors}, the two routines share the same underlying communication code, so the scaling performance of each is similar.

\subsection{Particle-Mesh Operations \label{sec:particles:particle_mesh}}
Many of the application codes that use AMReX's particle data structures implement some form of particle-mesh method, in which quantities are interpolated between the particle positions and the cells, edges, faces, or nodes of a structured mesh. Example use-cases include Lagrangian tracers (in which this interpolation is one-way only), models of drag forces between a fluid and colloidal particles, and electrostatic and electromagnetic Particle-in-Cell schemes.

To perform these operations in a distributed setting, some type of parallel communication is required. We opt to communicate the mesh data rather than the particle data for these operations, because the connectivity of the mesh cells is simpler and thus the operation can be made more efficient. However, this may not be the best choice when the particles are very sparse and/or the interpolation stencil is very wide. Depending on the type and order of interpolation employed, the mesh data involved in these interpolations requires some number of ghost cells to ensure that each grid has enough data to fully perform the interpolation for the particles inside it. For mesh-to-particle interpolation, we use \texttt{FillBoundary} to update the ghost cells on each grid prior to performing the interpolation. For particle-to-mesh (i.e. deposition), we use the analogous \texttt{SumBoundary} operation after performing a deposition, which adds the values from overlapping ghost cells back to the corresponding valid cell, ensuring that all the particle's contributions are accounted for. We provide generic, lambda-function based \texttt{ParticleToMesh} and \texttt{MeshToParticle} functions for performing general interpolations. Additionally, users can always write their own \texttt{MFIter} loops to implement these operations in a more customized way.

For particle-to-mesh interpolation, an additional factor is that race conditions need to be taken into account when multiple CPU or GPU threads are in use. Our approach to this depends on whether AMReX is compiled for a CPU or GPU platform. When built for CPUs, we use AMReX's tiling infrastructure to assign different tiles to different OpenMP threads. Each thread then deposits its particles into a thread-local buffer, for which no atomics are needed. We then perform a second reduction step where each tile's local buffer is atomically added to the \texttt{MultiFab} storing the result of the deposition. For GPUs, no local buffer is used; each thread performs the atomic writes directly into global memory. This strategy works surprisingly well for NVIDIA V100 GPUs such as those on Summit, provided that we periodically sort the particles on each grid to better exploit the GPU's memory hierarchy.

\begin{figure}[h]
    \centering
    \includegraphics[width=0.45\textwidth]{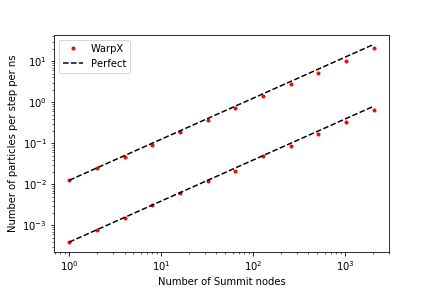}
    \caption{Weak scaling study of a WarpX uniform plasma benchmark on up to 2048 nodes on Summit. The $x$-axis is the number of nodes, while the $y$-axis displays a type of Figure-of-Merit: the total number of particles advanced per step per ns. Perfect weak scaling would be indicated by dark dashed lines. Results are shown for both CPU-only and GPU-accelerated runs. The CPU-only runs used all 42 cores per compute node, while the GPU-accelerated runs used all 6 GPUs per node. The overall speedup from using the accelerators is $\approx 30$ at all problem sizes.}
    \label{fig:warpx_weak_scaling}
\end{figure}

Figure \ref{fig:warpx_weak_scaling} shows the results of a weak scaling study on a uniform plasma benchmark from the AMReX-using code WarpX \citep{WarpX}. Of all AMReX users, WarpX places the most stress on the particle-mesh operations, since the deposition of current from macro-particles is often the dominant cost. This benchmark also stresses several parallel communication routines, including particle redistribution, \texttt{FillBoundary}, and \texttt{SumBoundary}. 

Additionally, the WarpX project has identified a plasma-acceleration run as its ECP KPP benchmark. At the time of this writing, the figure of merit from this benchmark measured on 4263 Summit nodes and extrapolated to the full machine is over 100 times larger than the baseline, which was measured using the original Warp code on Cori. 

\subsection{Particle-Particle operations}

Another common need in AMReX particle applications - for example, MFIX-Exa and the fluctuating hydrodynamics code DISCOS, is to pre-compute a neighbor list, or a set of possible collision partners for each particle over the next N time steps. AMReX provides algorithms for constructing these lists that run on both the CPU and the GPU based on a user-provided decision function that returns whether a pair of particles belong in each other's lists or not. These algorithms employ a spatial binning method similar to that used in Redistribution that avoids an $N^2$ search over the particles on a grid. Tools for iterating over neighbor lists are also provided; see Listing \ref{lst:nbors} for example code.

\begin{lstlisting}[language=C++,frame=single,basicstyle=\footnotesize,label={lst:nbors}, caption={Iterating over particles in a neighbor list}]
amrex::ParallelFor ( np, 
[=] AMREX_GPU_DEVICE (int i)
{
    ParticleType& p1 = parts[i];

    for (const auto& p2 : nlist.getNeighbors(i))
    {
        Real dx = p1.pos(0) - p2.pos(0);
        Real dy = p1.pos(1) - p2.pos(1);
        Real dz = p1.pos(2) - p2.pos(2);
        ...
    }
}
\end{lstlisting}

\subsection{Parallel Scan}
When running on GPUs, many particle operations can be expressed in terms of a parallel prefix sum, which computes a running tally of the values in an array. For example, a commonly-needed operation is to compute a permutation array that puts the particles on a grid into a bin-sorted order based on some user-specific bin size. This array can then be used to re-order the particles in memory, or it can be used directly to access physically nearby particles in an out-of-order fashion. To perform this operation in AMReX, we use a counting sort algorithm, which involves as one of its stages an exclusive prefix sum. Other operations that can be expressed in terms of a prefix sum include partitioning the particles in a grid, various stream compaction operations (filtering, scattering, gathering), and constructing neighbor lists.

AMReX provides an implementation of prefix sum based on \cite{Merrill2016SinglepassPP} that works with CUDA, HIP, and DPC++. Our API allows callables to be passed in that define arbitrary functions to be called when reading in and writing out data. See Listing \ref{lst:scan} for an example. This allows for the expression of, e.g., a partition function that avoids unnecessary memory traffic for temporary variables.

\begin{lstlisting}[language=C++,frame=single,basicstyle=\footnotesize,label={lst:scan}, caption={API for PrefixSum.}]
PrefixSum<T>(n,
[=] AMREX_GPU_DEVICE (int i) -> T 
{ return in[i]; },
[=] AMREX_GPU_DEVICE (int i, T const& x) 
{ out[i] = x; },
Type::exclusive);
\end{lstlisting}

Using this scan implementation, AMReX provides tools for performing the above binning and stream compaction operations. These tools are used internally in AMReX when, for example, redistributing particles or constructing neighbor lists. They are also used by AMReX applications, for example to implement DSMC-style collisions where particles in the same bin undergo random collisions, or to model a variety of particle creation processes in WarpX, such as ionization, pair production, and quantum synchrotron radiation.

\subsection{Dual grid approach for load balancing}
\label{sec:particles:dual_grid}
In AMReX-based applications that have both mesh data and particle data, the mesh work and particle work have very different requirements for load balancing.
Rather than using a combined work estimate to create the same grids for mesh and particle data, AMReX supplies the option to pursue a “dual grid” approach.
With this approach the mesh (\texttt{MultiFab}) and particle (\texttt{ParticleContainer}) data are allocated on different \texttt{BoxArray}s with different \texttt{DistributionMapping}s.
This enables separate load balancing strategies to be used for the mesh and particle work.  An MFiX-Exa simulation, for example, may have very dense and very dilute particle distributions within the same domain, while the dominant cost of the fluid evolution -- the linear solver for the projections -- is evenly distributed throughout the domain.
The cost of this strategy, of course, is the need to copy mesh data onto temporary \texttt{MultiFab}s defined on the particle \texttt{BoxArray}s when mesh-particle communication is required.
   
\section{Embedded boundary representation}

\begin{figure}[h]
    \centering
    \includegraphics[width=0.4\textwidth]{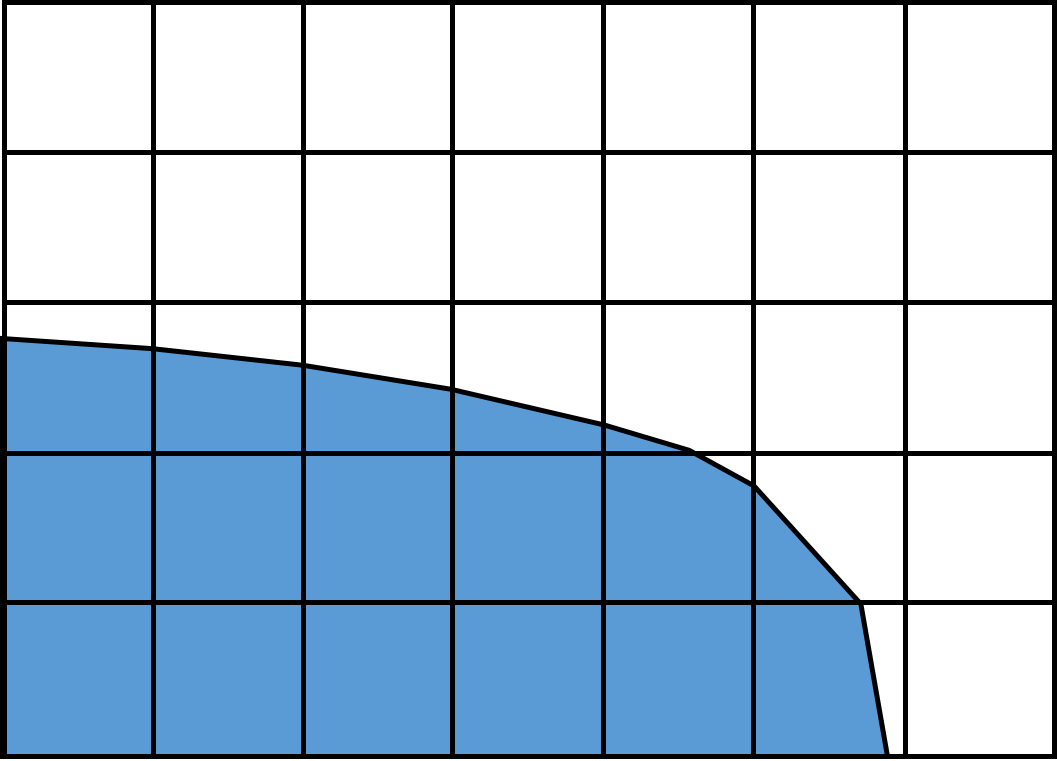}
    \caption{Sketch illustrating embedded boundary representation of geometry.}
    \label{fig:eb_sketch}
\end{figure}

AMReX supports complex geometries with an embedded boundary (EB)
approach that fits naturally within the block-structured AMR approach.  In
this approach, the problem geometry is represented as an interface that cuts
through a regular computational mesh as shown in Figure \ref{fig:eb_sketch}. Thus, although the computational domain has an irregular
shape, the underlying computational mesh is uniform at each level of the AMR 
hierarchy.  

\subsection{Data structures and algorithms}

When an embedded boundary is present in the domain, there are three types of cells: regular, cut and covered, as seen in Figure \ref{fig:eb_sketch}.  AMReX provides data structures for
accessing EB information, and supports various algorithms and solvers
for solving partial differential equations on AMR grids with complex
geometries.

The EB information is precomputed and stored in a distributed
database at the beginning of the calculation.  It is available for AMR
meshes at each level and for coarsened meshes in multigrid solvers.  The
information includes cell type, volume fraction, volume centroid, face area
fractions and face centroids.  For cut cells, the information also
includes the centroid, normal and area of the EB face.
Additionally, there is connectivity information between neighboring
cells.

The implementation of the EB data structures uses C\texttt{++}
\texttt{std::shared\_ptr}s. Regular data structures such as
\texttt{MulitFab}s and \texttt{FArrayBox}s have the option to store a
shared copy of the EB data.  One can query an \texttt{FArrayBox} to
find out if it has any cut or covered cells within a given
\texttt{Box}.  This allows the user to adapt an existing non-EB code
to support EB by adding support to handle regions with EB. Computations
with EB are usually not balanced in work load.  Regions with cut cells
are more expensive than regions with only regular cells, whereas the
completely covered regions have no work.  Applications can measure the
wallclock time spent in different regions and use that information for load
balancing when regridding is performed.

Numerical algorithms that use an embedded boundary typically use specialized stencils in cells containing and near the boundary; in particular they may pay special attention to the treatment of the solution in ``small cells" that can arise when the problem domain
is intersected with the mesh.
\commentout{AMReX provides support for accessing the geometric information
needed to construct EB-aware discretizations as well as support for specialized algorithms needed to support these types of discretizations.}

\commentout{In addition to the EB information, AMReX also provides help for common
operations in EB algorithms on AMR grids.}  For example, due to the
stability constraint for cut cells with very small volume fractions,
EB-aware algorithms often need to redistribute part of the update from
small cut cells to their neighbors.  Furthermore, to maintain
conservation when relevant, both the fluxes and redistribution corrections at AMR coarse/fine boundaries need to be
synchronized.  AMReX provides a class \texttt{EBFluxRegister} for
refluxing in solvers for  hyperbolic conservation laws.  AMReX also
provides EB-aware interpolation functions for data on different AMR levels.

\subsection{Geometry Generation}

\begin{figure}[h]
    \centering
    \includegraphics[width=0.45\textwidth]{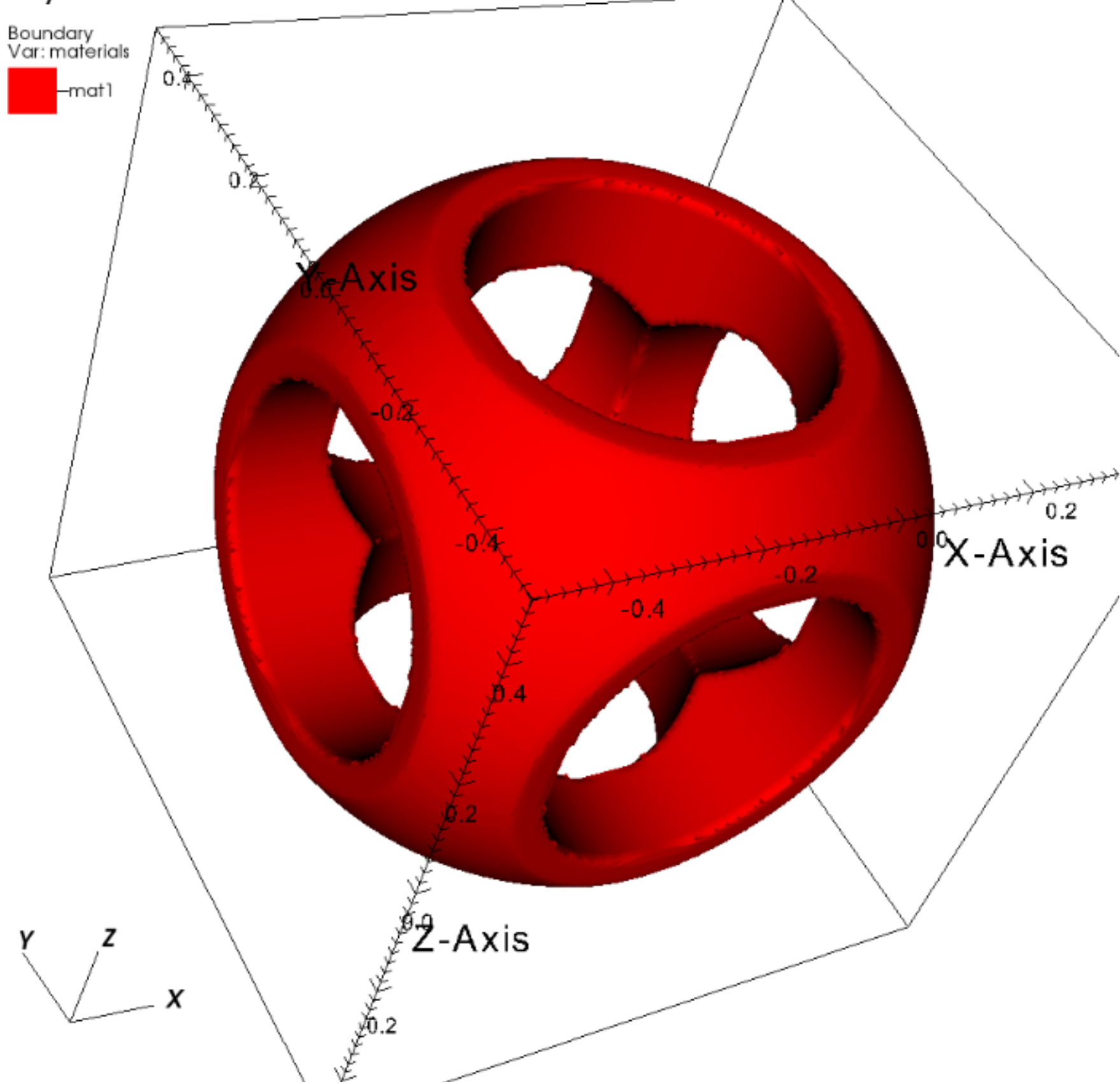}
    \caption{Example of complex geometry generated with CSG from simple shapes.}
    \label{fig:csg}
\end{figure}

AMReX provides an implicit function approach for generating the
geometry information.  In this approach, there is an implicit function
that describes the surface of the embedded object.  It returns a
positive value, a negative value or zero, for a given position inside
the body, inside the fluid, or on the boundary, respectively.
Implicit functions for various simple shapes such as boxes, cylinders,
spheres, etc., as well as a spline based approach, are provided.
Furthermore, basic operations in constructive solid geometry (CSG) such as
union, intersection and difference are used to combine objects
together.  Geometry transformations (e.g., rotation and translation)
can also be applied to these objects. Figure~\ref{fig:csg} shows an
example of geometry generated with CSG from simple shapes using the implicit
function defined in Listing~\ref{lst:csg}. Besides, the implicit function
based approach, an application code can also use its own approach to
generate geometry information and then store it in AMReX's EB database.

\begin{lstlisting}[language=C++,frame=single,basicstyle=\footnotesize,label={lst:csg},caption={Example of constructing a complex implicit function based on simple objects (sphere, box and cylinder) with transformations(intersection, union and difference).}]
  EB2::SphereIF sphere(0.5, {0.0,0.0,0.0}, false);
  EB2::BoxIF cube({-0.4,-0.4,-0.4}, {0.4,0.4,0.4}, false);
  auto cubesphere = EB2::makeIntersection(sphere, cube);
  EB2::CylinderIF cylinder_x(0.25, 0, {0.0,0.0,0.0}, false);
  EB2::CylinderIF cylinder_y(0.25, 1, {0.0,0.0,0.0}, false);
  EB2::CylinderIF cylinder_z(0.25, 2, {0.0,0.0,0.0}, false);
  auto three_cylinders = EB2::makeUnion(cylinder_x, cylinder_y, cylinder_z);
  auto csg = EB2::makeDifference(cubesphere, three_cylinders);
\end{lstlisting}

\subsection{Mesh pruning}

\begin{figure}[h]
    \centering
    \includegraphics[width=0.45\textwidth]{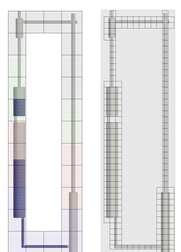}
    \caption{Mesh pruning of a Chemical Looping Reactor geometry. Shown here is the EB geometry and the individual grids at level 0 for a relatively coarse simulation.  Courtesy of MFiX-Exa team.}
    \label{fig:clr}
\end{figure}

When the simulation domain is contained by EB surfaces, ``mesh pruning" can reduce the memory required by eliminating fully covered grids at each level (i.e. grids that are fully outside of the region in which the solution will be computed) from the \texttt{BoxArray} at that level.   As long as the \texttt{MultiFabs} holding the solution are created after the \texttt{BoxArray} has been pruned, no data will be allocated in those regions.  This capability has been demonstrated for a simplified Chemical Looping Reactor (CLR) geometry by the MFiX-Exa project; see Figure~\ref{fig:clr} for an example of how mesh pruning becomes even more effective as the base mesh is resolved.

\subsection{Particle-Wall interactions}

In cases where particles interact with EB surfaces, AMReX provides the functionality to build a level set, with values defined at nodes, that represents the distance function from the EB surface.  Particle-wall collisions can then be computed based on the level set values at the nodes surrounding the cell holding the particle.  In some applications, such as MFiX-Exa, the level set can be defined at a finer level than the rest of the simulation in order to exploit a finer representation of the geometry for calculating collisions.

\section{Linear Solvers}

A key feature of a number of applications based on AMReX is that they require solution of one or more linear systems at each time step.   AMReX includes native single-level and multi-AMR-level geometric multigrid and Krylov (CG and BiCG) solvers for nodal or cell-centered data, as well as interfaces to external solvers such as those in hypre \citep{hypre-paper} and PETSc \citep{petsc-web-page}. The external solvers can be called at the finest multigrid level, or as a "bottom solve" where the native solver coarsens one or more levels, then calls the external solver to reduce the residual by a specified tolerance at that level.  

Discretizations include variable coefficient Poisson and Helmholtz operators as well as the full viscous tensor for fluid dynamics (requiring solution of a multi-component system). 
The AMReX multigrid solvers include aggregation (merging boxes at a level in the multigrid hierarchy to enable additional coarsening within the V-cycle) and consolidation (reducing  the  number  of  ranks to reduce communication costs at coarser multigrid levels) strategies to reduce total cost.

%Second, regardless of the specific solution procedure, the efficient solution of elliptic equations at scale requires attention to efficient global communication.  The third implication is that the effectiveness of general task scheduling approaches may be constrained by the synchronization points/barriers imposed by linear solvers within a time step.

The linear solvers in AMReX also support complex geometries represented using the
EB approach.  For cell-centered data, the stencil uses EB
information such as the boundary normal, and  employs a geometric coarsening strategy.
The nodal linear solver in AMReX is based on a
finite-element approach.  The construction of the stencil uses
various pre-computed moment integrals
(e.g., $\iiint x^\alpha y^\beta  z^\gamma \,dx\,dy\,dz$,  %and $\iiint y z^2 \,dx\,dy\,dz$), and
with an algebraic mulitgrid approach to coarsening the operator.

\begin{figure}[h]
    \centering
    \includegraphics[width=0.45\textwidth]{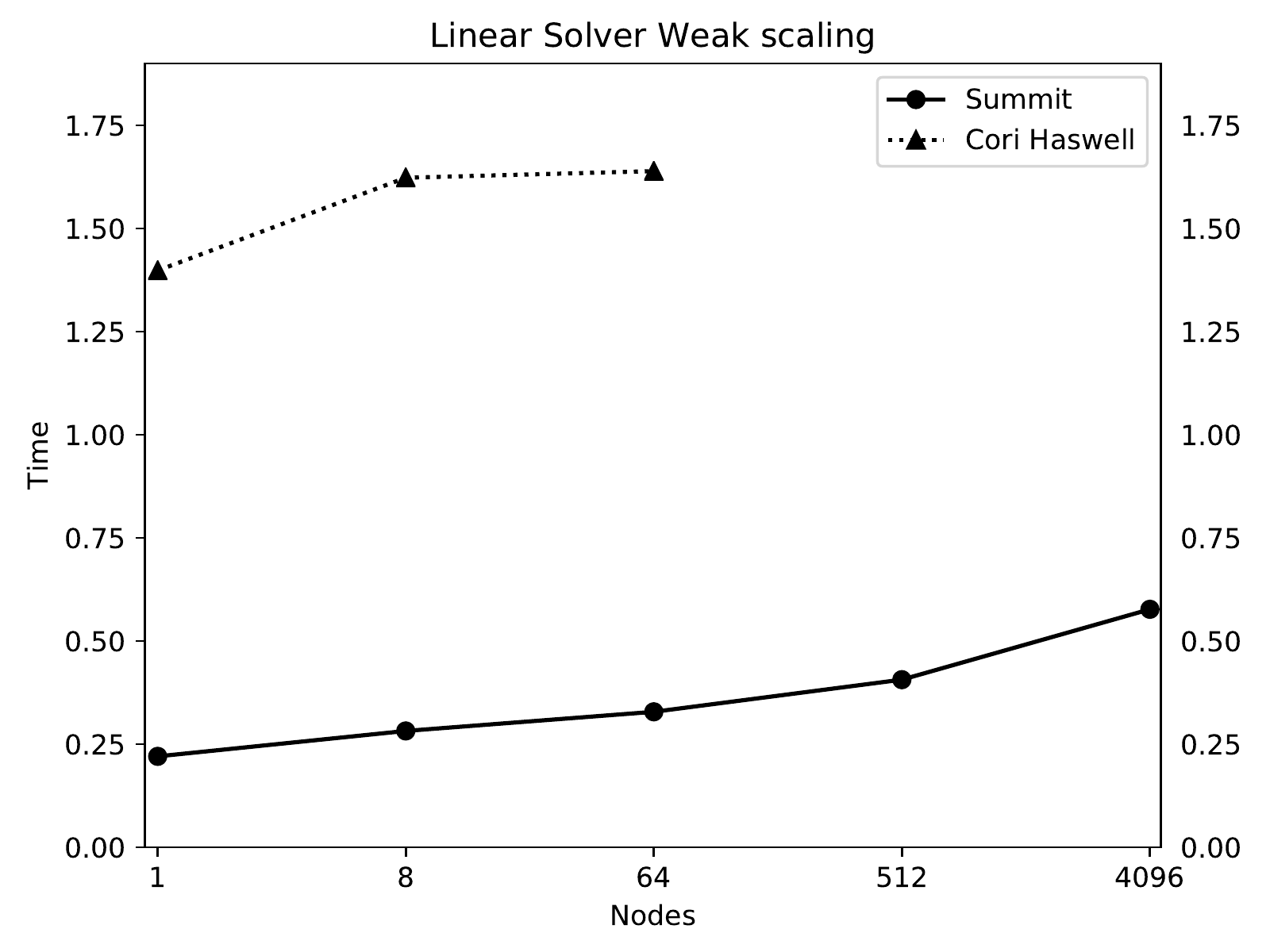}
    \caption{Weak scaling study of linear solver.  There are $256^3$
    cells per node.}
    \label{fig:solver}
\end{figure}

Figure~\ref{fig:solver} shows a weak scaling study of the
cell-centered linear solve for
\begin{equation}
a \alpha \phi - b \nabla \cdot (\beta \nabla \phi) = \mathrm{rhs},
\end{equation}
where $\phi$ is the unknown defined at cell centers, $a$ and $b$ are scalars,
$\alpha$ is defined at cell centers and may vary in space, and $\beta$ is a variable coefficient defined on
cell faces.  The calculations were performed on Summit with 6 GPUs per
node, and on Cori Haswell nodes with 32 CPU cores per node.  The weak
scaling tests had $256^3$ cells per node, and there were $4096^3$
cells for the 4096 nodes jobs on Summit.  On a single node, the GPU
run was 6 times faster than the CPU run.  The scaling behavior is consistent
with the characterization that multigrid solvers are communication heavy
especially because of the floating point operation performance of
GPUs.

\section{I/O}
AMReX provides a native file format for plotfiles that store the solution at a given time step for visualization.  Plotfiles use a well-defined format that is supported by a variety of third-party tools for visualization.  AMReX provides functions that directly write plotfiles from mesh and particle objects.  Mesh and particle plotfiles are written independently for improved flexibility and performance.  AMReX also provides users the option to use HDF5 for data analysis and visualization. 

Writing a plotfile requires coordination between MPI ranks to prevent overwhelming the I/O system with too many simultaneous writes to the file system.  AMReX has implemented multiple output methodologies to provide efficient I/O across a variety of applications and simulations. A static output pattern prints in a pre-determined pattern that eliminates unnecessary overhead, which is useful for well-balanced or small simulations.
%\MarginPar{what is the static write?   how is what is use determined?  }
% (Exactly as described: writing in a pre-determined or guessed pattern, instead of the task-like pattern of dynamic. Changed the wording a little.) (Generally, you use static for simple things and change if I/O is getting expensive or you're minimizing time for some run. I don't know if that's easy to fit in here (or worth it).
A dynamic output pattern improves write efficiency for complex cases by assigning ranks to coordinate the I/O in a task-like fashion.  Finally, asynchronous output assigns the writing to a background thread, allowing the computation to continue uninterrupted while the write is completed on a stored copy of the data.  The I/O output methodology and other standard features, such as the number of simultaneous writes, can be chosen through run-time and compile-time flags.

\begin{figure}[h]
    \centering
    \includegraphics[width=0.45\textwidth]{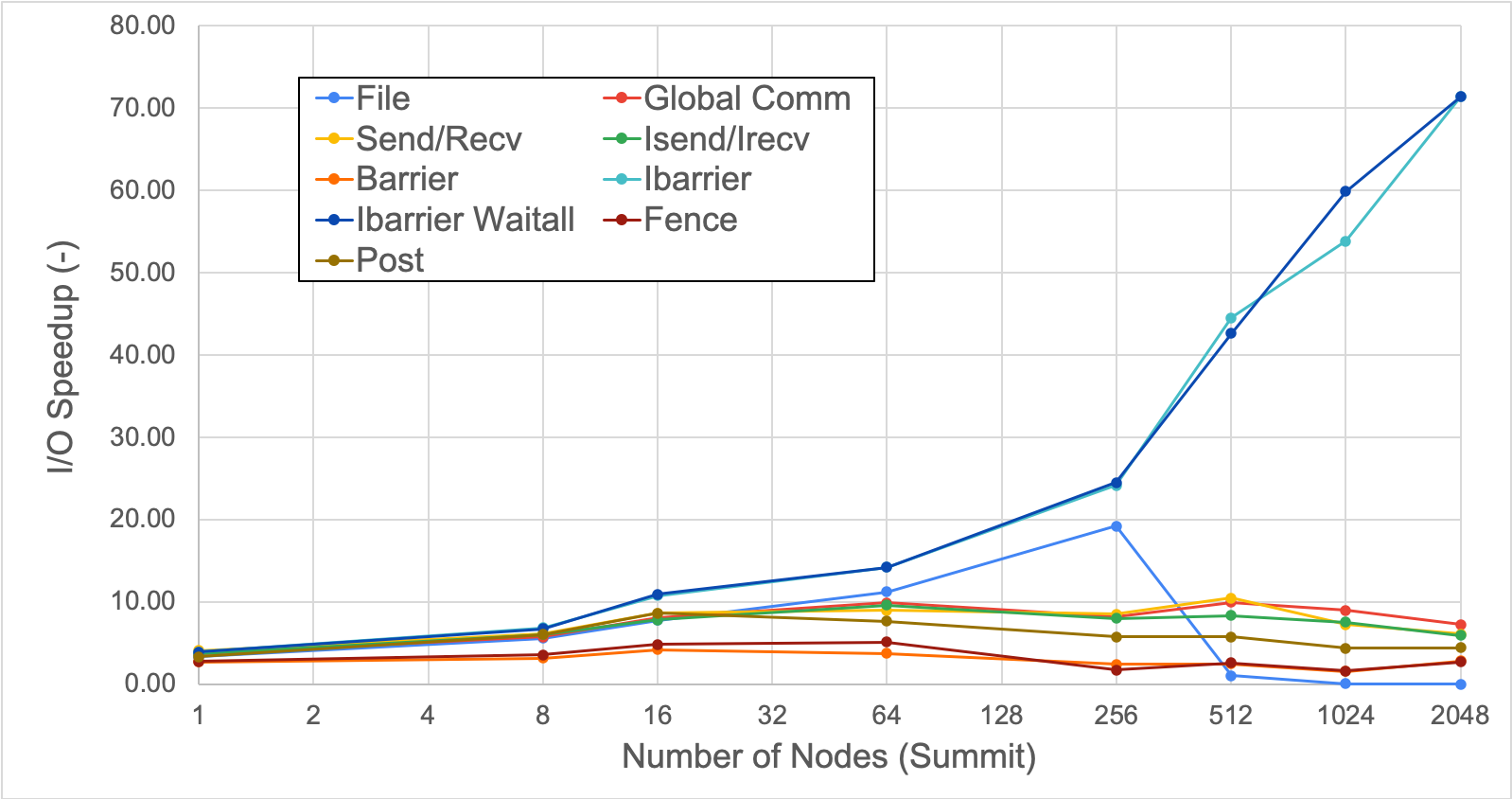}
    \caption{Weak scaling study of possible MPI implementations for Async I/O. Async I/O requires a zero-size message between I/O ranks.  A variety of possibilities were tested while AMReX Reduction calls \texttt{amrex::Min()} and \texttt{amrex::Max()} are called continuously on another thread to test \texttt{MPI\_THREAD\_MULTIPLE} overhead.  The results show only \texttt{MPI\_IBarrier} yields scalable results. \texttt{MPI\_IBarrier}'s performance is equivalent to less than 2\% overhead on full-scale Summit.}
    \label{fig:async_io}
\end{figure}

Asynchronous I/O is currently the targeted I/O method for exascale systems because it is a portable methodology that substantially reduces I/O impact on total run-time.  AMReX's native Async I/O has reduced write time by a factor of around 80 on full-scale Summit simulations, as shown in Figure \ref{fig:async_io}.  However, asynchronous I/O requires a scalable \texttt{MPI\_THREAD\_MULTIPLE} implementation to achieve the best results.  Scalable Async I/O requires one thread passing zero-size messages while another thread performs it's own communication with as little overhead as possible. On systems where \texttt{MPI\_THREAD\_MULTIPLE} implementations scale poorly, asynchronous I/O may not be the best choice.

AMReX also provides the infrastructure, examples and documentation to allow applications to write checkpoint files for simulation restarts.  Each application requires a unique set of data to restart successfully that is not known to AMReX, such as run-time determined constants or any pre-calculated data sets that cannot be easily reconstructed.  Therefore, each application has to build their own checkpoint system.  However, numerous examples and helpful functions are provided to help facilitate its construction.

\subsection{Visualization}

The AMReX plotfile format is supported by Paraview, Visit and yt.  
\commentout{Support from these visualization development teams is ongoing.}  
\commentout{ As of version 5.8, Paraview has the ability to read and visualize AMReX's native mesh and particle data, rather than needing to load them in separate instances.}  AMReX also supports a native visualization tool, \cite{AMRvis}, which users and developers find useful for testing, debugging and other quick-and-dirty visualization needs.  
The AMREX team is working with the ALPINE and Sensei teams, part of the ECP Software Technologies, to provide support for {\it in situ} data analysis and visualization; see, e.g. \cite{Biswas:2020}.  Preliminary interfaces to both exist in the AMReX git repo.
\commentout{and the ExaSky and MFiX-Exa projects are currently exploring this functionality.}

\commentout{
\section{Task Iteration Support -- fork-join, async ?? (1/2 page)}
\OLD{
Fork-Join Parallelism In many multi-physics applications, within each time step there are different physical processes that can be advanced independently.  AMReX includes support for higher-level asynchronous task description and execution based on a fork-join approach.  To enable fork-join task parallelism, a simple programming interface is provided for developers to express their coarse-grained tasks and the data those tasks require.  The runtime mechanisms for parallel task forking, handling data migration, and task output redirection is also provided. 
}
}

\section{Software Engineering}

AMReX is fully open source; source code, documentation, tutorials, and a gallery of examples are all hosted on Github (https://amrex-codes.github.io/amrex/).  AMReX supports building applications with both GNUMake and CMake and can also be built as a library.  Development takes place using a fork-and-pull request workflow, with stable tags released monthly.  Github CI is used to verify builds with a variety of compilers on Windows, Linux and Mac for each pull request.  In addition, a suite of regression tests are performed nightly, both on AMReX and on selected AMReX-based applications.  These tests cover both CPU architectures and NVIDIA GPUs.

In addition, AMReX has been part of the xSDK releases starting in 2018, and is available through spack.  AMReX is also one of the software packages featured in the Argonne Training Program for Extreme-Scale Computing (ATPESC); see  https://xsdk-project.github.io/MathPackagesTraining2020/lessons/amrex for the AMReX tutorial presented in 2020. 

\commentout{
\subsection{Relationships to ECP Software Technologies}

Several AMReX-based applications require integration of stiff systems, typically representing chemical or nuclear reaction networks.   Specifically, the Nyx and Pele projects use the CVODE package supplied by SUNDIALS.   
In addition, the AMReX and SUNDIALS teams have worked together to establish interoperability in the form of AMReX applications exploiting SUNDIALS integrators at the MultiFab rather than single-cell level; a tutorial demonstrating the functionality is available at https://github.com/AMReX-Codes/ATPESC-codes.}

%\MarginPar{Something about PETSC and hypre here . . .mentioned in linear solvers}

%\MarginPar{vis stuff talked about in io section . . reiterate here?}

%Working with HDF5 (this is mentioned in IO section)

%Finally, the AMREX team works with the ALPINE project to determine the best ways for ALPINE and SENSEI to support the AMReX-based application projects.  Preliminary interfaces to both exist in the AMReX git repo and the ExaSky and MFiX-Exa projects are currently exploring this functionality; see, e.g. \cite{Biswas:2020}.

\subsection{Profiling and Debugging}

Profiling is a crucial part of developing and maintaining a successful scientific code, such as AMReX and its applications.  The AMReX community uses a wide variety of compatible profilers, including VTune, CrayPat, ARM Forge, Amrvis and the Nsight suite of tools.  AMReX includes its own lightweight instrumentation-based profiling tool, TinyProfiler.  TinyProfiler consists of a few simple macros to insert scoped timers throughout the application.  At the end of a simulation, TinyProfiler reports the total time spent in each region, the number of times it was called and the variation across MPI ranks.  This tool is on by default in many AMReX applications and includes NVTX markers to allow instrumentation when using Nsight.  

AMReX's make system includes a debug mode with a variety of useful features for AMReX developers.  Debug mode includes array bound checking, detailed built-in assertions, initialization of floating point numbers to NaNs, signal handling (e.g. floating point exceptions) and backtrace capability for crashes.  This combination of features has proven extremely helpful for catching common bugs and reducing development time for AMReX users.

\section{Conclusions and Future Work}

%summary

%\begin{itemize}
%    \item higher / mixed dimension
%    \item curvilnear / multiblock
%    \item links to CAD
%    \item code generation
%\end{itemize}

In this paper we have discussed the main features of the AMReX framework for the development of block-structured AMR algorithms.  In addition to core capability to support data defined on a hierarchical mesh, AMReX also provides support for different particle and particle/mesh algorithms and for an embedded boundary representation of complex problem geometry.  We have also discussed linear solvers that are needed to support implicit discretizations and AMReX's I/O capabilities that are used to write data for data analysis and visualization and for checkpoint / restart.

A cross-cutting feature in all AMReX development is the need to provide performance portability.  With the push toward exascale we have seen the emergence of a variety of different architectures with disparate hardware capabilities and programming models.  AMReX incorporates an abstraction layer that isolates applications from the details of a particular architecture without sacrificing performance.  The framework also provides custom support for parallel reductions and memory management.
We have shown examples that illustrate that AMReX provides scalable performant implementations for a range of tasks typically found in multiphysics applications.

There are a number of additional capabilities that could be added to AMReX.  Currently, the framework is restricted to one, two or three spatial dimensions.  Extending the methodology to higher dimensions would be relatively straightforward.  In a similar vein, the framework could also be generalized to handle mixed dimensions in which different components of the physical model are solved in different dimensions.

The framework currently assumes grid cells are rectilinear with an embedded boundary representation for complex geometries.  A useful generalization to increase geometric capabilities would be to incorporate curvilinear coordinates in which the grid is logically rectangular but the individual cells can be generic hexahedra.  This type of capability could be further generalized to a mapped multiblock or overset mesh model in which multiple curvilinear meshes are coupled in a single simulation.  

Finally, one potentially exciting area is interfacing AMReX with a code generation system.  The basic idea here is to use a symbolic manipulation system that can generate complilable AMReX code directly from equations.  This type of capability is particularly useful for problems that can be stated concisely mathematically but expand dramatically when translated into code.  We have used a prototype code generation system for discretization of general relativity and a spectral representation of Boltzmann's equation.  The ability to describe complex problems concisely has potential to further increase the range of applications that can use AMReX.

\commentout{
\subsection{ECP Vendor Interaction  - JB} 

The AMReX vendor liaison continues to interact with a number of vendors, reading PathForward milestones and participating in PathForward reviews. 
The liaison regularly provides feedback geared towards improving the  sustained  performance  that  future  systems  will  deliver  on  AMReX-based  and  similar  applications.
Summaries of the architectural trends and implications have been discussed with ECCN/RSNDA-cleared personnel within AMReX in order to ensure that AMReX software development is in line with trends in architecture and system software.

\subsection{industry, ...}
 
\subsection{non-ECP}
\subsubsection{Brandon Runnels}
\subsubsection{Knut, Curistec, ...}

\REPORT{
\section{2019 ANNUAL REPORT}
\subsection{intro}

The goal of this project is to develop a new framework,  AMReX, to support the development of block-structured AMR algorithms for solving systems of partial differential equations on exascale architectures.Block-structured AMR provides the basis for the temporal and spatial discretization strategy for a large number of applications relevant to DOE. 

AMR reduces the computational cost and memory footprint compared to a uniform mesh while preserving the local descriptions of different physical processes in complex multi-physics algorithms.  Fundamental to block-structured AMR algorithms is a hierarchical representation of the solution at multiple levels of resolution.
At each level of refinement, the solution is defined on the union of data containers at that resolution, each of which represents the solution over a logically rectangular subregion of the domain.  
Solution strategies vary from level-by-level approaches (with or without subcycling in time) with multilevel synchronization to full-hierarchy approaches, and any combination thereof.  AMReX provides data containers and iterators that understand the underlying hierarchical parallelism for field variables on a mesh, particle data and embedded boundary (cut cell) representations of complex geometries.  Both particles and embedded boundary representations introduce additional irregularity and complexity in the way data is stored and operated on,requiring special attention in the presence of the dynamically changing hierarchical mesh structure and AMR time stepping approaches.The AMReX team is working closely with application partners to ensure that the software meets their requirements.  The team is also working closely with a number of ST projects to take advantage of new tools that are being developed.  Finally, the team is engaged in a dialogue with hardware vendors to provide them with information about adaptive mesh algorithms and provide feedback on the impact of hardware design decisions on AMR application

\subsection{Algorithms and Software Objectives}

Within this broad framework, the applications supported represent a wide range of multi-physics problems that couple a variety of different processes and have different computational requirements.  Many of these processes are described by systems of partial differential equations that are discretized on a mesh.  Discretization strategies for these processes often use either explicit discretizations that express updates in terms of the local state or implicit discretizations that require solution of linear systems.  In some cases, the problem includes stiff systems of ordinary differential equations that represent single-point processes such as chemical kinetics or nucleosynthesis.  Many AMReX-based applications also utilize Lagrangian particles to representsome aspect of the solution.  Particles play a variety of different roles in different applications, ranging from passively advected quantities used for analysis to playing the dominant role in the overall dynamics.  Several applications have a requirement for complex geometries.  For these types of applications, the team is developingan efficient embedded boundaries (EB) representation, in which the solid boundaries are represented as aninterface that cuts through a regular adaptive mesh on which the fluid variables are defined.

At the core of the AMReX software is a flexible set of data structures that can be used to represent block-structured mesh data in a distributed memory environment.  Operations supported on these datastructures include iterators for operations at a level, a communications layer to handle ghost cell exchangeand data distribution and tools for synchronization between levels.  The iterators support logical tiling withOpenMP on CPU-based architectures, as well as kernel launching and effective use of managed memory onhybrid CPU/GPU systems.This basic framework includes native geometric multigrid solvers, with support for solving systems arisingfrom embedded boundary discretizations, as well as interfaces to external solvers such as hypre and PETSc.

The team provides an interface to SUNDIALS for integration of stiff ODEs.  On top of this core functionalitythe team is also developing a rich and flexible set of tools for treating Lagrangian particles.  These tools allowfor different representations of particle data (Structure-of-Arrays (SoA) versus Array-of-Structures (AoS)),particle communications and support for particle algorithms in an AMR context.  

For embedded boundaryrepresentations of complex geometry, AMReX provides support for the necessary geometric information.

Additionally, AMReX provides tools for regridding operations and load balancing; a fast I/O layer for writing checkpoint, restart, and visualization/analysis data; and a rich set of native profiling tools.

  While many core discretizations, such as standard second-order and some fourth-order spatial and temporal discretizations, are provided within the AMReX framewor kfor the convenience of users and developers, AMReX allows application developers sufficient access to the underlying data structures to allow them to implement and optimize new discretizations as well

Communication
AMReX provides MPI communication routines that operate directly on GPU memory buffers, so that unnecessary device-to-host and host-to-device copies are not triggered on machines, such as OLCF’s Summit,that support GPUDirect.  For particles,  this requires operations such as sorting,  searching,  and streamc ompaction that work efficiently on the device.  The current approach to this problem uses Thrust, a parallelExascale Computing Project (ECP)148ECP-XX-XXXX
algorithms library maintained by NVIDIA and distributed as part of the CUDA Toolkit.  As the AMReX-basedapplications move more of their functionality onto GPUs, the team will work in conjunction with them to assess which components of the algorithm are most effectively offloaded to the GPU and which components stay on the CPU without negatively impacting overall performance.  
This raises the possibility of executing on the CPU and GPU in parallel and or executing different algorithm components on different GPUs.  This will potentially require some redefinition of the underlying algorithms as well as software infrastructure to support the implementation.  The team’s fork-join approach provides a suitable framework that could be generalized to handle these types of algorithms.

Performance characterization. Predicting the performance of complex multi-physics AMR applications a priori is challenging at best.  The team has developed a sophisticated set of profiling tools for performance characterization of AMReX-basedapplications.  The toolkit includes measurement of both computation and communication.  The software can be used to obtain anything from a broad overview of performance to detailed measurements localized to a specific portion of the algorithm.  In addition to a brief summary report, the tools create a database during execution that can be queried using a graphical interface to extract detailed information.  This overall performance measurement system is augmented with performance modeling methodology that enables the user to construct a model of execution including both compute and communication to understand performance behavior and how it depends on machine characteristics.  This type of performance model enables the user to evaluate the potential impact of algorithm changes on performance prior to full implementation.
}
}

\begin{acks}
%\MarginPar{official ECP acknowledgement}
This research was supported by the Exascale Computing Project (17-SC-20-SC), a collaborative effort of the U.S. Department of Energy Office of Science and the National Nuclear Security Administration.
%\MarginPar{LBL}
This work was supported by the
U.S.~Department of Energy, Office of Science,
Office of Advanced Scientific Computing Research,
Exascale Computing Project under contract DE-AC02-05CH11231.
%\MarginPar{NERSC}
This research used resources of the National Energy Research Scientific Computing Center, a DOE Office of Science User Facility supported by the Office of Science of the U.S. Department of Energy under Contract No. DE-AC02-05CH11231; 
%\MarginPar{OLCF}
and resources of the Oak Ridge Leadership Computing Facility at the Oak Ridge National Laboratory, which is supported by the Office of Science of the U.S. Department of Energy under Contract No. DE-AC05-00OR22725.
\end{acks}

\bibliographystyle{SageH}
\bibliography{exascale.bib}

\begin{biog}

\vspace{\baselineskip}
\noindent
{\it Weiqun Zhang} is a Computer Systems Engineer at Lawrence  Berkeley National Laboratory.  He is interested in high-performance computing, computational physics, and programming in general.  Currently, he works on the AMReX software framework and WarpX, an advanced electromagnetic Particle-in-Cell code.

\vspace{\baselineskip}
\noindent
{\it Andrew Myers} is a Computer Systems Engineer in the Center for Computational Sciences and Engineering at Lawrence Berkeley National Laboratory. His work focuses on scalable particle methods for emerging architectures in the context of adaptive mesh refinement. He is an active developer of the AMReX framework and of the electromagnetic Particle-in-Cell code WarpX.

\vspace{\baselineskip}
\noindent
{\it Kevin Gott} is an application performance specialist at NERSC.  His primary focus is supporting the AMReX Co-Design Center as it develops performance portable solutions for next generation Exascale supercomputers.  His interests include C++ programming, portable GPU code design, MPI performance and advanced parallel programming techniques.

\vspace{\baselineskip}
\noindent
{\it Ann Almgren} is a Senior Scientist at Lawrence Berkeley National Laboratory, and the Group Lead of LBL's Center for Computational Sciences and Engineering. Her primary research interest is in
computational algorithms for solving PDE’s in a variety of application areas. Her current projects include the development and implementation of new multiphysics algorithms in high-resolution adaptive mesh
codes that are designed for the latest hybrid architectures.
She is a Fellow of the Society of Industrial and Applied Mathematics and the Deputy Director of the AMReX Co-Design Center.

\vspace{\baselineskip}
\noindent
{\it John Bell} is a Senior Scientist at Lawrence Berkeley National Laboratory.   His research focuses on the development and analysis of numerical methods for partial differential equations arising in science and engineering. He is a Fellow of the Society of Industrial and Applied Mathematics, a member of the National Academy of Sciences, and the Director of the AMReX Co-Design Center..

\end{biog}
\end{document}